\documentclass[12pt,eqno,epsf]{article}
\usepackage{amsmath,amssymb,graphicx,slashbox}
\numberwithin{equation}{section}

\def\ignore#1{{}}

\newcounter{sxn}

\newcounter{axn}

\date{}

\newdimen\mybaselineskip
\renewcommand{\thefootnote}{\arabic{footnote}}

\newcommand{\beeq}{\begin{equation}}
\newcommand{\eneq}{\end{equation}}
\newcommand{\beqn}{\begin{eqnarray}}
\newcommand{\eeqn}{\end{eqnarray}}

\newcommand{\alp}{\alpha}
\newcommand{\bt}{\beta}
\newcommand{\gm}{\gamma}

\newcommand{\dlt}{\delta}

\newcommand{\ep}{\epsilon}
\newcommand{\tht}{\theta}

\newcommand{\lmd}{\lambda}
\newcommand{\Lmd}{\Lambda}
\newcommand{\sgm}{\sigma}
\newcommand{\Sgm}{\Sigma}

\newcommand{\omg}{\omega}
\newcommand{\Omg}{\Omega}
\newcommand{\dalp}{\dot{\alpha}}
\newcommand{\dbt}{\dot{\beta}}

\newcommand{\ualp}{\underline{\alp}}
\newcommand{\ubt}{\underline{\bt}}

\newcommand{\be}{\begin{equation}}
\newcommand{\ee}{\end{equation}}
\newcommand{\bea}{\begin{eqnarray}}
\newcommand{\eea}{\end{eqnarray}}
\newcommand{\eql}{\!\!\!&=\!\!\!&}

\newcommand{\defa}{\!\!\!&\equiv\!\!\!&}

\newcommand{\toa}{\!\!\!&\to\!\!\!&}

\newcommand{\exch}{\leftrightarrow}

\newcommand{\tl}[1]{\tilde{#1}}
\newcommand{\bdm}[1]{{\mbox{\boldmath $#1$}}}

\newcommand{\diag}{{\rm diag}}
\newcommand{\der}{\partial}
\newcommand{\dr}{\!\!d}
\newcommand{\hc}{{\rm h.c.}}
\newcommand{\ie}{{i.e.}}
\newcommand{\id}{\mbox{\boldmath $1$}}

\newcommand{\udl}[1]{\underline{#1}}

\newcommand{\vev}[1]{\langle #1 \rangle}

\newcommand{\brkt}[1]{\left( #1 \right)}
\newcommand{\brc}[1]{\left\{ #1 \right\}}
\newcommand{\sbk}[1]{\left[ #1 \right]}
\newcommand{\abs}[1]{\left| #1 \right|}

\renewcommand{\Re}{{\rm Re}\,}
\renewcommand{\Im}{{\rm Im}\,}

\newcommand{\cA}{{\cal A}}

\newcommand{\cL}{{\cal L}}
\newcommand{\cM}{{\cal M}}
\newcommand{\cN}{{\cal N}}
\newcommand{\cO}{{\cal O}}

\newcommand{\cQ}{{\cal Q}}

\newcommand{\cV}{{\cal V}}

\newcommand{\cW}{{\cal W}}

\newcommand{\bE}{{\mathbb E}}

\newcommand{\bH}{{\mathbb H}}

\newcommand{\bT}{{\mathbb T}}
\newcommand{\bV}{{\mathbb V}}

\newcommand{\cPT}{{\cal P}_{\rm T}}
\newcommand{\suU}{\mbox{SU(2)}_{\mbox{\scriptsize\bf U}}}

\begin{document}
\thispagestyle{empty}

\baselineskip=12pt


\begin{flushright}
KEK-TH-1848 \\
WU-HEP-15-14
\end{flushright}

\baselineskip=35pt plus 1pt minus 1pt

\vskip 1.5cm

\begin{center}
{\LARGE\bf $\bdm{\cN=1}$ superfield description of \\
six-dimensional supergravity}

\vspace{1.5cm}
\baselineskip=20pt plus 1pt minus 1pt

\normalsize

{\large\bf Hiroyuki Abe,}${}^1\!${\def\thefootnote{\fnsymbol{footnote}}
\footnote[1]{E-mail address: abe@waseda.jp}}
{\large\bf Yutaka Sakamura}${}^{2,3}\!${\def\thefootnote{\fnsymbol{footnote}}
\footnote[2]{E-mail address: sakamura@post.kek.jp}} 
{\large\bf and Yusuke Yamada}${}^1\!${\def\thefootnote{\fnsymbol{footnote}}
\footnote[3]{E-mail address: yuusuke-yamada@asagi.waseda.jp}}

\vskip 1.0em

${}^1${\small\it Department of Physics, Waseda University, \\ 
Tokyo 169-8555, Japan}

\vskip 1.0em

${}^2${\small\it KEK Theory Center, Institute of Particle and Nuclear Studies, 
KEK, \\ Tsukuba, Ibaraki 305-0801, Japan} \\ \vspace{1mm}
${}^3${\small\it Department of Particles and Nuclear Physics, \\
SOKENDAI (The Graduate University for Advanced Studies), \\
Tsukuba, Ibaraki 305-0801, Japan}

\end{center}

\vskip 1.0cm
\baselineskip=20pt plus 1pt minus 1pt

\begin{abstract}
We express the action of six-dimensional supergravity 
in terms of four-dimensional $\cN=1$ superfields,  
focusing on the moduli dependence of the action. 
The gauge invariance of the action in the tensor-vector sector is realized 
in a quite nontrivial manner, and it determines the moduli dependence of the action. 
The resultant moduli dependence is intricate, especially on the shape modulus. 
Our result is reduced to the known superfield actions 
of six-dimensional global SUSY theories and of five-dimensional supergravity 
by replacing the moduli superfields with their background values 
and by performing the dimensional reduction, respectively. 
\end{abstract}

\newpage

\section{Introduction}
Higher dimensional supergravity (SUGRA) theories provide interesting setups 
for supersymmetric (SUSY) models with extra dimensions, 
and are also regarded as effective theories of the superstring theory in some cases. 
For the purpose of analyzing SUSY extra-dimensional models, 
the $\cN=1$ superfield description of the action is 
quite useful~\cite{Marcus:1983wb}-\cite{Sakamura:2012bj}.\footnote{
``$\cN=1$'' denotes SUSY with four supercharges in this paper. 
}
It makes the derivation of four-dimensional (4D) effective theories transparent 
since the Kaluza-Klein mode expansion can be performed 
keeping the $\cN=1$ superspace structure. 
It also expresses the SUGRA action compactly, 
and allows us to work in general setups. 
In the global SUSY case, the $\cN=1$ superfield description of 
SUSY Yang-Mills theories from five to ten dimensions 
are provided in Ref.~\cite{ArkaniHamed:2001tb}. 
However, we have to work in the context of SUGRA 
in order to treat the moduli, which are dynamical degrees of freedom 
corresponding to the ``volume'' or the ``shape'' of the compactified 
internal space. 
Such moduli often play important roles when we construct  
phenomenologically viable models. 
We also need to discuss the stabilization of the moduli 
to some finite values to obtain consistent extra-dimensional models. 

Five-dimensional (5D) SUGRA provides the simplest setup for SUSY extra-dimensional models. 
The general action can be obtained 
by the superconformal formulation~\cite{Zucker:1999ej}-\cite{Kugo:2002vc}. 
Based on this formulation, 5D SUGRA action with arbitrary numbers of 
hyper and vector multiplets has been expressed 
in terms of $\cN=1$ superfields~\cite{Paccetti:2004ri,Abe:2004ar}. 
We have derived 4D effective theories of various 5D SUGRA models, 
and discussed their phenomenology~\cite{Abe:2006eg}-\cite{Sakamura:2014aja}. 

The next simplest case is six-dimensional (6D) SUGRA~\cite{Nishino:1984gk,Salam:1984cj}. 
This has the smallest even extra-dimensions, 
and we can introduce magnetic flux that penetrates the compact space 
as a background. 
The shape modulus newly appears in addition to the volume modulus. 
These ingredients widen the possibility of model-building. 
Besides, we can also consider 6D SUGRA as a toy model of 
ten-dimensional superstring theories. 
With these reasons, 6D SUGRA is intriguing subject to investigate. 
As mentioned above, the $\cN=1$ superfield description is useful to discuss it, 
as was provided in Ref.~\cite{ArkaniHamed:2001tb} in the global SUSY case. 
However, 6D action in Ref.~\cite{ArkaniHamed:2001tb} 
cannot be promoted to SUGRA straightforwardly. 
As discussed in Refs.~\cite{Bergshoeff:1985mz,Coomans:2011ih}, 
the off-shell description of 6D SUGRA necessarily contains a tensor multiplet, 
which was not introduced in Ref.~\cite{ArkaniHamed:2001tb}. 
It contains a self-dual antisymmetric tensor~$B^+_{MN}$ ($M,N=0,1,\cdots,5$), 
and the 6D superconformal Weyl multiplet 
contains an anti-self-dual tensor~$T_{MNL}^-$. 
In general, the (anti-)self-dual condition is an obstacle to 
the Lagrangian formulation, similar to that for type IIB SUGRA. 
Fortunately, we can evade this difficulty in 6D SUGRA. 
By combining $T_{MNL}^-$ with the field strength~$F_{MNL}^+\equiv\der_{[M}B^+_{NL]}$, 
we can define a new Weyl multiplet~\footnote{
This is called the ``Weyl 2 multiplet'' in Ref.~\cite{Coomans:2011ih}, 
and the ``type-II Weyl multiplet'' in Ref.~\cite{Linch:2012zh}.  
} 
that contains an unconstrained tensor~$B_{MN}$. 
This new tensor field couples to the vector multiplets~\cite{Bergshoeff:1985mz,Coomans:2011ih}. 
Therefore we need to know how the tensor and the vector multiplets couple to each other 
in the $\cN=1$ superfield language. 

In our previous work~\cite{Abe:2015bqa}, we derived the $\cN=1$ superfield description of 
the tensor-vector couplings in 6D global SUSY theories, 
which is derived from the invariant action~\cite{Linch:2012zh} 
in the projective superspace~\cite{Karlhede:1984vr,Lindstrom:1987ks,Lindstrom:1989ne}. 
In this case, the tensor multiplet must be treated as external fields 
because we do not have the Weyl multiplet that contains $T_{MNL}^-$, 
and only have the constrained one~$B_{MN}^+$. 
In this paper, we extend our result in Ref.~\cite{Abe:2015bqa} to SUGRA. 
Since Ref.~\cite{Linch:2012zh} provides the projective superspace formulation 
of 6D SUGRA, we can in principle obtain its $\cN=1$ superfield description by integrating out 
half of the Grassmannian coordinates, as we did in the global SUSY case~\cite{Abe:2015bqa}. 
However, the procedure is not so straightforward as that in the global SUSY case 
because we need to separately treat the 4D part and the extra-dimensional part of 
the gravity sector that has a complicated structure in the projective superspace. 
Hence we adopt another strategy. 
We first identify the moduli superfields that originate from the extra-dimensional 
components of the 6D Weyl multiplet. 
Then, we insert them into the action  
in the global SUSY case under the following requirements. 
\begin{enumerate}
\item The action is reduced to 
the global SUSY one if the moduli superfields are replaced with their background values. 
\item It is consistent with the component field expression of the action. 
\item It is invariant under the supergauge transformations. 
\end{enumerate}
The superfield action is uniquely determined by these requirements. 
As a nontrivial check, we show that our result reproduces  
the known superfield action of 5D SUGRA obtained in Refs.~\cite{Paccetti:2004ri,Abe:2004ar} 
after the dimensional reduction. 

The paper is organized as follows. 
In the next section, we give a brief review of the superfield description 
of 6D global SUSY theories. 
In Sec.~\ref{6DSUGRAaction}, we promote it to the local SUSY case, 
and identify the desired superfield action of 6D SUGRA. 
In Sec.~\ref{consistency_checks}, we explicitly show the gauge invariance of our result  
and the consistency with the known 5D SUGRA action through the dimensional reduction. 
Sec.~\ref{summary} is devoted to the summary. 
We also collect some formulae and their derivation in the appendices. 

\section{6D Global SUSY theory}
Throughout the paper, we take the metric convention as $\eta_{MN}=\diag(-1,1,1,1,1,1)$, 
and follow the notation of Ref.~\cite{Wess:1992cp} 
for the 2-component spinors. 

\subsection{Invariant action}
We consider 6D (1,0) SUSY theories. 
The spacetime coordinates~$x^M$ ($M=0,1,\cdots,5$) are decomposed into 
the 4D ones~$x^\mu$ ($\mu=0,1,2,3$) and the extra dimensional 
ones~$x^m$ ($m=4,5$). 
Before discussing 6D SUGRA, let us begin with its global SUSY limit. 
In this case, it is convenient to use 
the complex coordinates~$z\equiv\frac{s}{2}(x^4-ix^5)$ 
($s\equiv e^{-\frac{\pi}{4}i}$) and its complex conjugate~$\bar{z}$,\footnote{
The definition of $z$ is different from that of Ref.~\cite{Abe:2015bqa}. 
As we will see in the next section, this choice is convenient 
for the promotion to SUGRA. 
} 
instead of $x^m$. 
Originally, the $\cN=1$ description of the action is provided in Ref.~\cite{ArkaniHamed:2001tb}. 
For simplicity, we will consider Abelian gauge theories. 
The field content consists of hypermultiplets~$\bH^A$ ($A=1,2,\cdots$)
and vector multiplets~$\bV^I$ ($I=1,2,\cdots$).  
They are decomposed into $\cN=1$ superfields as 
\bea
 \bH^A \eql (H^{2A-1},H^{2A}), \;\;\;\;\;
 \bV^I = (V^I,\Sgm^I), 
\eea
where $V^I$ is an $\cN=1$ real vector superfield, while the others 
are chiral superfields. 
By using these $\cN=1$ superfields, 
we can construct 6D global SUSY action as~\cite{ArkaniHamed:2001tb}
\bea
 S_{\rm global} \eql \int\dr^6x\;\brkt{\cL_{\rm V}+\cL_{\rm H}}, \nonumber\\
 \cL_{\rm V} \defa \brc{\int\dr^2\tht\;\frac{f_{IJ}}{2}\cW^I\cW^J+\hc} 
 \nonumber\\
 &&+\int\dr^4\tht\;f_{IJ}
 \brc{4(\bar{\der}V^I-\bar{\Sgm}^I)(\der V^J-\Sgm^J)
 -2\bar{\der}V^I\der V^J}, \nonumber\\
 \cL_{\rm H} \defa \int\dr^4\tht\;
 2\brc{(H_{\rm odd}^\dagger e^V H_{\rm odd}
 +H_{\rm even}^\dagger e^{-V}H_{\rm even}}
 \nonumber\\
 &&-\sbk{\int\dr^2\tht\;\brc{
 H_{\rm odd}^t\brkt{\der-\Sgm}H_{\rm even}
 -H_{\rm even}^t\brkt{\der+\Sgm}H_{\rm odd}}+\hc},  \label{cL:global:1}
\eea
where $\der\equiv\der_z=\bar{s}(\der_4+i\der_5)=s^{-1}\der_4-s\der_5$, and 
$H_{\rm odd}$ and $H_{\rm even}$ are column vectors that consist of 
$H^{2A-1}$ and $H^{2A}$, respectively. 
The contracted indices~$I$ and $J$ are understood as being summed, and 
\be
 \cW^I_\alp \equiv -\frac{1}{4}\bar{D}^2D_\alp V^I 
\ee
is the gauge-invariant field strength superfield. 
The coefficients~$f_{IJ}$ are real constants and $f_{IJ}=f_{JI}$. 
The superfields without the indices~$V$ and $\Sgm$ are defined as 
\be
 V \equiv t_I V^I, \;\;\;\;\;
 \Sgm \equiv t_I\Sgm^I, 
\ee
where $t_I$ ($I=1,2,\cdots$) are generators for the corresponding Abelian gauge groups. 
The Lagrangian~(\ref{cL:global:1}) is invariant under the following 
(super)gauge transformation. 
\bea
 V^I \toa V^I+\Lmd^I+\bar{\Lmd}^I, \;\;\;\;\;
 \Sgm^I \to \Sgm^I+\der\Lmd^I, \nonumber\\
 H_{\rm odd} \toa e^{-\Lmd}H_{\rm odd}, \;\;\;\;\;
 H_{\rm even} \to e^\Lmd H_{\rm even},  \label{gg_trf}
\eea
where the transformation parameter~$\Lmd^I$ is a chiral superfield. 

Unfortunately, (\ref{cL:global:1}) cannot be promoted to SUGRA straightforwardly. 
As mentioned in the introduction, a tensor multiplet~$\bT=\{B_{MN}^+,\cdots\}$ 
is necessary to describe 6D SUGRA. 
Thus we need to extend (\ref{cL:global:1}) including $\bT$ 
in order to promote the action to the SUGRA one. 
This extension was provided in our previous work~\cite{Abe:2015bqa}, 
which is directly derived from the invariant action 
in the 6D projective superspace~\cite{Linch:2012zh}.  
We have to note that the tensor multiplet~$\bT$ cannot be off-shell 
in the global SUSY case~\cite{Sokatchev:1988aa}. 
We found that it is expressed by two $\cN=1$ superfields, \ie, 
a real linear superfield~$\Phi_T$ and a chiral spinor superfield~$\cW_{T\alp}$, 
which are subject to the constraints:
\bea
 D^\alp\cW_{T\alp} \eql -2\bar{\der}\Phi_T, \nonumber\\
 \bar{D}^2D_\alp\Phi_T \eql -4\der\cW_{T\alp}.  \label{constraints}
\eea
From these relations, we obtain
\be
 \brkt{\Box_4+\der\bar{\der}}\Phi_T = \brkt{\Box_4+\der\bar{\der}}\cW_{T\alp} = 0, 
\ee
where $\Box_4\equiv \eta^{\mu\nu}\der_\mu\der_\nu$. 
We have used that $\cPT\Phi_T=\Phi_T$ and 
$\bar{D}^2D^2\cW_{T\alp}=16\Box_4\cW_{T\alp}$, 
where $\cPT\equiv -\bar{D}_{\dalp}D^2\bar{D}^{\dalp}/(8\Box_4)$. 
Namely, $\Phi_T$ and $\cW_{T\alp}$ are on-shell, and thus 
should be treated as external superfields. 
Using these superfields, $\cL_{\rm V}$ in (\ref{cL:global:1}) is extended to
\bea
 \cL_{\rm VT} \eql -\sbk{\int\dr^2\tht\;f_{IJ}\brc{2\Sgm^I\cW^J\cW_T
 +\frac{1}{4}\bar{D}^2\brkt{\Phi_TD^\alp V^I\cW_\alp^J
 +\der V^ID^\alp V^J\cW_{T\alp}}}+\hc} \nonumber\\
 &&+\int\dr^4\tht\;2f_{IJ}\Phi_T\brc{V^I\brkt{\Box_4\cPT+\der\bar{\der}}V^J
 +2(\bar{\der}V^I-\bar{\Sgm}^I)(\der V^J-\Sgm^J)}. 
\eea
For later convenience, we rewrite this Lagrangian as 
\bea
 \cL_{\rm VT} \eql \int\dr^4\tht\;f_{IJ}\left[
 \brc{-2\Sgm^ID^\alp V^J\cW_{T\alp}
 +\frac{1}{2}\brkt{\der V^I D^\alp V^J
 -\der D^\alp V^IV^J}\cW_{T\alp}+\hc} \right.\nonumber\\
 &&\hspace{20mm}
 +\Phi_T\left\{D^\alp V^I\cW_\alp^J+\bar{D}_{\dalp}V^I\bar{\cW}^{J\dalp}
 +V^ID^\alp\cW_\alp^J \right.\nonumber\\
 &&\hspace{30mm}\left.\left.
 +4(\bar{\der}V^I-\bar{\Sgm}^I)(\der V^J-\Sgm^J)-2\bar{\der}V^I\der V^J\right\}\right], 
 \label{cL:global:2}
\eea
where we have dropped total derivatives and used the first constraint 
in (\ref{constraints}). 
As we have shown in Ref.~\cite{Abe:2015bqa}, this Lagrangian is invariant 
under the gauge transformation~(\ref{gg_trf})~\footnote{
The tensor multiplet~$(\Phi_T,\cW_{T\alp})$ is invariant under the gauge transformation. 
}
up to total derivatives, 
and reduces to (\ref{cL:global:1}) in the limit of $\Phi_T=1$ and $\cW_{T\alp}=0$, 
which corresponds to the case where the tensor multiplet is absent. 

The superfields~$\Phi_T$ and $\cW_{T\alp}$ are expressed as
\bea
 \Phi_T \eql -2iD^\alp\bar{D}^2Y_{\alp}+2i\bar{D}_{\dalp}D^2\bar{Y}^{\dalp}, 
 \nonumber\\
 \cW_{T\alp} \eql i\bar{D}^2\brkt{D_\alp\bar{X}+4\bar{\der}Y_\alp}, 
 \label{def:PhiW_T}
\eea
where $X$ and $Y_{\alp}$ are complex superfields 
that are related through
\be
 \bar{D}^2\brkt{D_\alp X+4\der Y_\alp} = 0.  \label{rel:XYZ}
\ee
This relation indicates that $Y_\alp$ cannot be 
a general superfield. 
The first constraint in (\ref{constraints}) 
is automatically satisfied if (\ref{rel:XYZ}) is satisfied. 
Thus, independent constraints are (\ref{rel:XYZ}) 
and the second constraint in (\ref{constraints}). 
Note that $\Phi_T$ and $\cW_{T\alp}$ are the field strength superfields 
of the ``gauge potentials''~$X$ and $Y_\alp$, and are invariant under 
\be
 X \to X+\der V_G-\Sgm_G, \;\;\;\;\;
 Y_\alp \to Y_\alp-\frac{1}{4}D_\alp V_G,  \label{ggtrf:XY}
\ee
where the transformation parameters~$V_G$ and $\Sgm_G$ 
are $\cN=1$ real vector and chiral superfields, and 
form a 6D vector multiplet. 
The transformation~(\ref{ggtrf:XY}) is the SUSY extension of the gauge transformation: 
$B^+_{MN}\to B^+_{MN}+\der_M\lmd_N-\der_N\lmd_M$ 
($\lmd_M$: real transformation parameter). 

Here we decompose $X$ as
\be
 X = s^{-1}X_4-s X_5, 
\ee
where $X_4$ and $X_5$ are real superfields. 
Then the second equation in (\ref{def:PhiW_T}) and (\ref{rel:XYZ}) are rewritten as
\be
 \cW_{T\alp} = \bar{D}^2\brc{s^{-1}D_\alp X_4+s D_\alp X_5
 +4\brkt{s^{-1}\der_4+s\der_5}Y_\alp},  \label{def:cW_T:global}
\ee
and 
\be
 \bar{D}^2\brkt{s^{-1}D_\alp X_4-s D_\alp X_5+4\der Y_\alp} = 0. 
 \label{rel:XYZ:2}
\ee
Using the constraint~(\ref{rel:XYZ:2}), $\cW_{T\alp}$ is also expressed as
\bea
 \cW_{T\alp} \eql 2s^{-1}\bar{D}^2\brkt{D_\alp X_4+4\der_4 Y_\alp}
 = s^{-1}\cW_{4\alp}+8s^{-1}\der_4\bar{D}^2Y_\alp \nonumber\\
 \eql 2s\bar{D}^2\brkt{D_\alp X_5+4\der_5 Y_\alp} 
 = s\cW_{5\alp}+8s\der_5\bar{D}^2Y_\alp, \label{def:cW_T:2}
\eea
where
\be
 \cW_{4\alp} \equiv 2\bar{D}^2D_\alp X_4, \;\;\;\;\;
 \cW_{5\alp} \equiv 2\bar{D}^2D_\alp X_5.  \label{def:cW_45}
\ee
Thus, the tensor multiplet~$\bT$ is described by two constrained 
superfields~$X_4$ (or $X_5$) and $Y_\alp$.

\ignore{
The second constraint can also be relaxed by adding the following terms 
to the Lagrangian. 
\be
 \cL_{\rm LM} = \int\dr^4\tht\;2i\tl{Y}^\alp
 \brkt{\bar{D}^2D_\alp\Phi_T+4\der\cW_{T\alp}}+\hc,  \label{def:cL_LM}
\ee
where a general superfield~$\tl{Y}^\alp$ is the Lagrange multiplier. 
This can be understood as the invariant Lagrangian constructed 
from two tensor multiplets 
if we regard $\tl{Y}^\alp$ as a superfield for 
the second tensor multiplet~\cite{Abe:2015bqa}. 
}

\subsection{Components of superfields}
Each $\cN=1$ superfield has the following components. 
Here we focus on the bosonic fields, for simplicity. 

Hyperscalars~$(\cA_i^{2A-1},\cA_i^{2A})$ in $\bH^A$, 
where $i=1,2$ is an $\suU$-doublet-index,\footnote{
$\suU$ is an automorphism of 6D superconformal algebra (see Appendix~\ref{SCalgebras}). 
}  
are embedded into $H^{2A-1}$ and $H^{2A}$ as
\be
 H^{2A-1} = \cA_2^{2A-1}+\cO(\tht), \;\;\;\;\;
 H^{2A} = \cA_2^{2A}+\cO(\tht).  \label{compid:hyper:1}
\ee

A 6D vector field~$A_M^I$ in $\bV^I$ is embedded into $V^I$ and $\Sgm^I$ as 
\be
 V^I = -(\tht\sgm^\mu\bar{\tht})A_\mu+\cO(\tht^3), \;\;\;\;\;
 \Sgm^I = \brkt{s^{-1}A_4-s A_5}+\cO(\tht). \label{compid:vector:1}
\ee

A 6D tensor field~$B_{MN}^+$ and its scalar partner~$\sgm$ in $\bT$  
are embedded into $\Phi_T$ and $\cW_{T\alp}$ as
\bea
 \Phi_T \eql \sgm+(\tht\sgm^\mu\bar{\tht})
 \ep_{\mu\nu\rho\lmd}\der^\nu B^{+\rho\lmd}
 -\frac{1}{4}\tht^2\bar{\tht}^2\Box_4\sgm+\cdots,  \nonumber\\
 \cW_{T\alp} \eql \tht_\alp\bar{\der}\sgm
 +(\sgm^{\mu\nu}\tht)_\alp\brkt{\bar{\der}B_{\mu\nu}^++\der_\mu C_\nu-\der_\nu C_\mu}
 +\cdots, 
 \label{comp:PhicW_T}
\eea
where $C_\mu\equiv -i\brkt{s^{-1}B_{\mu 4}^++s B_{\mu 5}^+}$, 
and $B_{MN}^+$ satisfies the self-dual condition:
\bea
 \ep_{\mu\nu\rho\lmd}\der^\nu B^{+\rho\lmd} 
 \eql -2\brc{\der_\mu B_{45}^+-\Im(\der C_\mu)}, \nonumber\\
 \bar{\der}B_{\mu\nu}^++\der_\mu C_\nu-\der_\nu C_\mu 
 \eql \frac{i}{2}\ep_{\mu\nu\rho\lmd}
 \brkt{\bar{\der}B^{+\rho\lmd}+\der^\rho C^\lmd-\der^\lmd C^\rho}. 
\eea
The expressions in (\ref{comp:PhicW_T}) are realized 
when $X$ and $Y_\alp$ have the following components: 
\bea
 X \eql \frac{1}{4}(\tht\sgm^\mu\bar{\tht})\bar{C}_\mu
 -\frac{1}{8}\tht^2\bar{\tht}^2\brkt{B_{45}^++\frac{i}{2}\sgm}+\cdots, 
 \nonumber\\
 Y_\alp \eql \frac{1}{16}\tht_\alp\bar{\tht}^2\brkt{B_{45}^++\frac{i}{2}\sgm}
 +\frac{i}{16}(\sgm^{\mu\nu}\tht)_\alp\bar{\tht}^2 B^+_{\mu\nu}+\cdots. 
 \label{comp:XY}
\eea
where $\bar{C}_\mu=s^{-1}B_{\mu 4}^+-sB_{\mu 5}^+$. 
The $B_{45}^+$-dependence is determined from the transformation property 
under (\ref{ggtrf:XY}).

\section{Extension to 6D SUGRA} \label{6DSUGRAaction}
Now we extend the action in the previous section to the local SUSY case. 
Since we are interested in the moduli-dependence of the action, 
we focus on $e_m^{\;\;\udl{n}}$ ($m,n=4,5$) among the sechsbein~$e_M^{\;\;\udl{N}}$, 
and treat the other components as a background,\footnote{
The fluctuation modes of the 4D gravity multiplet can be easily 
taken into account by promoting the $d^4\tht$- and $d^2\tht$-integrals 
to the D-term and the F-term action formulae~\cite{Kugo:1982cu}, respectively,  
in the superconformal formulation of 4D SUGRA~\cite{Kaku:1977rk,Kaku:1977pa,Kaku:1978ea}. \label{FN:4Dgravity}
} 
\ie, $e_\mu^{\;\;\udl{\nu}}=\dlt_\mu^{\;\;\nu}$ and 
$e_\mu^{\;\;\udl{n}}=e_m^{\;\;\udl{\nu}}=0$.
Therefore, we do not discriminate the curved index~$\mu$ from the flat index~$\udl{\mu}$ 
for the 4D part in the following.

\subsection{Moduli superfields} \label{moduli_sf}
First we identify the $\cN=1$ superfields constructed from 
the extra-dimensional components of 
the 6D Weyl multiplet~$\bE=(e_M^{\;\;\udl{N}},\Psi^i_{M\udl{\alp}},V_M^{ij},\cdots)$ 
(see Appendix~\ref{trf:Weyl}.).  
Notice that if a complex scalar~$\cA$ is the lowest component 
of a chiral superfield, it transforms under consecutive SUSY transformations as
\be
 \dlt_\ep\dlt_\eta\cA = 2i(\eta\sgm^\mu\bar{\ep})\der_\mu\cA+\cdots, 
\ee
and if a real scalar~$\phi$ is the lowest component of a real general superfield, 
it transforms as
\be
 \dlt_\ep\dlt_\eta\phi = i(\eta\sgm^\mu\bar{\ep}-\ep\sgm^\mu\bar{\eta})\der_\mu\phi
 +\cdots, 
\ee
where the 2-component spinors~$\ep_\alp$ and $\eta_\alp$ are the transformation parameters, 
and the ellipses denote terms involving other fields. 
In order to identify combinations of $e_m^{\;\;\udl{n}}$ that belong to 
$\cN=1$ superfields, we focus on the $\cN=1$ SUSY transformations at linearized level 
in the fluctuations~$\tl{e}_m^{\;\;\udl{n}}$. 
Then, from (\ref{SUSYtrf:Weyl}), we obtain
\be
 \dlt_\ep\dlt_\eta u = \frac{1}{2\vev{e^{(2)}}}(\eta\sgm^\mu\bar{\ep})
 \vev{\cM}\der_\mu u+{\rm c.c.}+\cdots, 
\ee
where $e^{(2)}\equiv\det(e_m^{\;\;\udl{n}})=e_4^{\;\;\udl{4}}e_5^{\;\;\udl{5}}
-e_4^{\;\;\udl{5}}e_5^{\;\;\udl{4}}$ and 
$u\equiv (\tl{e}_4^{\;\;\udl{4}},\tl{e}_4^{\;\;\udl{5}},\tl{e}_5^{\;\;\udl{4}},
\tl{e}_5^{\;\;\udl{5}})^t$. 
The matrix~$\cM$ is defined as 
\be
 \cM \equiv 
 \begin{pmatrix} \cM_{11} & \cM_{12} 
 & -E_4e_4^{\;\;\udl{4}} & -E_4 e_4^{\;\;\udl{5}} \\
 -i\cM_{11} & -i\cM_{12} & iE_4e_4^{\;\;\udl{4}} & iE_4e_4^{\;\;\udl{5}} \\
 E_5e_5^{\;\;\udl{4}} & E_5e_5^{\;\;\udl{5}} & \cM_{33} & \cM_{34} \\
 -iE_5e_5^{\;\;\udl{4}} & -iE_5e_5^{\;\;\udl{5}} & 
 -i\cM_{33} & -i\cM_{34} \end{pmatrix},  
\ee
where $E_m\equiv e_m^{\;\;\udl{4}}+ie_m^{\;\;\udl{5}}$, and 
\bea
 \cM_{11} \defa 2E_5e_4^{\;\;\udl{4}}-E_4e_5^{\;\;\udl{4}}, \;\;\;\;\;
 \cM_{12} \equiv 2E_5e_4^{\;\;\udl{5}}-E_4e_5^{\;\;\udl{5}}, \nonumber\\
 \cM_{33} \defa E_5e_4^{\;\;\udl{4}}-2E_4e_5^{\;\;\udl{4}}, \;\;\;\;\;
 \cM_{34} \equiv E_5e_4^{\;\;\udl{5}}-2E_4e_5^{\;\;\udl{5}}. 
\eea
There are three eigenvectors~$v_a$ ($a=\pm,0$) that satisfy 
$v_a\vev{\cM}=\lmd_av_a$ and $v_a\vev{\cM}^*=\lmd'_av_a$ simultaneously 
($\lmd_a$, $\lmd'_a$: eigenvalues). 
\bea
 (\lmd_-,\lmd'_-) = (0,-4i\vev{e^{(2)}}) & : & 
 v_- = \brkt{\vev{\bar{E}_5},-i\vev{\bar{E}_5},-\vev{\bar{E}_4},i\vev{\bar{E}_4}}, 
 \nonumber\\
 (\lmd_0,\lmd'_0) = (2i\vev{e^{(2)}},-2i\vev{e^{(2)}}) & : & 
 v_0 = \brkt{\vev{e_5^{\;\;\udl{5}}},-\vev{e_5^{\;\;\udl{4}}},
 -\vev{e_4^{\;\;\udl{5}}},\vev{e_4^{\;\;\udl{4}}}}, \nonumber\\
 (\lmd_+,\lmd'_+) = (4i\vev{e^{(2)}},0) & : & 
 v_+ = \brkt{\vev{E_5},i\vev{E_5},-\vev{E_4},-i\vev{E_4}}. 
\eea
Thus, we obtain 
\bea
 \dlt_\ep\dlt_\eta(v_-\cdot u) \eql 
 -2i(\ep\sgm^\mu\bar{\eta})\der_\mu(v_-\cdot u)+\cdots, \nonumber\\
 \dlt_\ep\dlt_\eta(v_0\cdot u) \eql 
 i(\eta\sgm^\mu\bar{\ep}-\ep\sgm^\mu\bar{\eta})\der_\mu(v_0\cdot u)+\cdots, 
 \nonumber\\
 \dlt_\ep\dlt_\eta(v_+\cdot u) \eql 
 2i(\eta\sgm^\mu\bar{\ep})\der_\mu(v_+\cdot u)+\cdots. 
\eea
Therefore, we infer that $v_+\cdot u=\vev{E_5}\tl{E}_4-\vev{E_4}\tl{E}_5$ 
is the lowest component of a chiral superfield, and 
$v_0\cdot u=\vev{e_5^{\;\;\udl{5}}}\tl{e}_4^{\;\;\udl{4}}
-\vev{e_5^{\;\;\udl{4}}}\tl{e}_4^{\;\;\udl{5}}
-\vev{e_4^{\;\;\udl{5}}}\tl{e}_5^{\;\;\udl{4}}
+\vev{e_4^{\;\;\udl{4}}}\tl{e}_5^{\;\;\udl{5}}$ 
is the lowest component of a real general superfield.\footnote{
$v_-\cdot u=(v_+\cdot u)^*$ is the lowest component of an anti-chiral superfield.
} 
Note that $v_+\cdot u$ and $v_0\cdot u$ are the linear parts 
of $E_4/E_5$ and $e^{(2)}$ in the fluctuations, respectively. 
In fact, we can show that 
\bea
 (\dlt_\ep\dlt_\eta-\dlt_\eta\dlt_\ep)\frac{E_4}{E_5} 
 \eql 2i\brkt{\eta\sgm^\mu\bar{\ep}-\ep\sgm^\mu\bar{\eta}}\der_\mu\brkt{\frac{E_4}{E_5}}, 
 \nonumber\\
 (\dlt_\ep\dlt_\eta-\dlt_\eta\dlt_\ep)e^{(2)} 
 \eql 2i\brkt{\eta\sgm^\mu\bar{\ep}-\ep\sgm^\mu\bar{\eta}}\der_\mu e^{(2)}, 
\eea
at the full order in the fluctuation. 
Thus the correct SUSY algebra is realized on them, and they can be 
the components of the superfields. 
Namely, we find that the extra-dimensional components of 
the 6D Weyl multiplet~$\bE$ form a chiral superfield,\footnote{
When $E_4/E_5$ is the lowest component of a chiral superfield, 
so is $(E_4/E_5)^p$ ($p$ : real number). 
We choose $p=1/2$ just for convenience. 
}  
\be
 S_E = \sqrt{\frac{E_4}{E_5}}+\cO(\tht),  \label{comp:S_E}
\ee
and a real general superfield, 
\be
 V_E = e^{(2)}+\cO(\tht).  \label{comp:V_E}
\ee
In the superconformal formulation of 4D 
SUGRA~\cite{Kugo:1982cu}-\cite{Kaku:1978ea},
each superconformal multiplet is characterized 
by the Weyl weight~$w$ and the chiral weight~$n$, 
which are the charges of the dilatation and the automorphism~$U(1)_A$ 
of the superconformal algebra, respectively. 
From (\ref{id:cQ_A}), we can see that $E_m$ ($m=4,5$) have $(w,n)=(-1,-1)$. 
Thus, noting that $e^{(2)}=\Im(\bar{E}_4E_5)$, we find that
$S_E$ and $V_E$ have $(w,n)=(0,0)$ and $(-2,0)$, respectively.  
This is consistent with the fact 
that they are a chiral and a real general superfields~\cite{Kugo:1982cu}.  
From their forms of the lowest components, we can see that $V_E$ and $S_E$ 
correspond to the ``volume'' and the ``shape'' of the compact space. 
 
In the following, we identify how these superfields appear 
in the 6D SUGRA action. 
We construct the action in such a way that 
it is reduced to the global SUSY one if 
the moduli superfields~$V_E$ and $S_E$ are replaced with 
constant values~1 and $s=e^{-\frac{\pi}{4}i}$, respectively. 
These values correspond to the background values of the case that 
$\vev{e_4^{\;\;\udl{4}}}=\vev{e_5^{\;\;\udl{5}}}=1$ 
and $\vev{e_4^{\;\;\udl{5}}}=\vev{e_5^{\;\;\udl{4}}}=0$.

\subsection{Hypermultiplet sector}
Here we extend $\cL_{\rm H}$ in (\ref{cL:global:1}) to the SUGRA version. 
In this case, we need to introduce the $n_C$ compensator 
hypermultiplets in addition to the $n_P$ physical ones. 
Thus, besides the dependence on $S_E$ and $V_E$, the Lagrangian in this sector 
is written as 
\bea
 \cL_{\rm H} \eql -\int\dr^4\tht\;2\brkt{H_{\rm odd}^\dagger\tl{d}e^V H_{\rm odd}
 +H_{\rm even}^\dagger\tl{d}e^{-V}H_{\rm even}} \nonumber\\
 &&+\sbk{\int\dr^2\tht\;\brc{H_{\rm odd}^t\tl{d}\brkt{\der-\Sgm}H_{\rm even}
 -H_{\rm even}^t\tl{d}(\der+\Sgm)H_{\rm odd}}+\hc}, 
 \label{cL_hyp:2}
\eea
where $\tl{d}=\diag(\id_{n_C},-\id_{n_P})$ is the metric for 
the space spanned by the hyperscalars, and discriminates the compensators  
from the physical ones. 

Now we consider the moduli dependence of the Lagrangian. 
Since $V_E$ cannot appear in the chiral superspace, 
$H_{\rm odd}$ and $H_{\rm even}$ must have $w=n=3/2$. 
However, the 6D hyperscalars~$\cA_i^{2A-1}$ and $\cA_i^{2A}$ have $w=n=2$.  
Hence the component identification in (\ref{compid:hyper:1}) must be modified. 
Since we have to keep the condition~$w=n$ for a chiral superfield, 
we need to adjust the weights by using $E_m=e_m^{\;\;\udl{4}}+ie_m^{\;\;\udl{5}}$ 
($m=4,5$) that has $w=n=-1$. 
We find that (\ref{compid:hyper:1}) should be modified as
\bea
 H^{2A-1} \eql E_4^p E_5^{1/2-p}\cA_2^{2A-1}+\cO(\tht), \nonumber\\
 H^{2A} \eql E_4^q E_5^{1/2-q}\cA_2^{2A}+\cO(\tht), 
\eea
where $p$ and $q$ are arbitrary real numbers. 
We can always set $p=q=1/4$ by redefining the above chiral superfields as 
$S_E^{1/2-2p}H^{2A-1}\to H^{2A-1}$ 
and $S_E^{1/2-2q}H^{2A}\to H^{2A}$. 
Hence, in the following, 
we identify the lowest components of these chiral superfields as
\bea
 H^{2A-1} \eql (E_4E_5)^{1/4}\cA_2^{2A-1}+\cO(\tht), \nonumber\\
 H^{2A} \eql (E_4E_5)^{1/4}\cA_2^{2A}+\cO(\tht).  \label{comp:Phis}
\eea

Next we promote the derivative~$\der$ to the SUGRA version~$\der_E$ 
that depends on $S_E$. 
(This is independent of $V_E$ because it cannot appear in the chiral superspace.) 
In order to reproduce the correct 6D kinetic terms for the hyperscalars 
after eliminating the F-terms of $H_{\rm odd,even}$, 
the lowest component of $\der_E$ should be proportional to $\der_{\udl{4}}+i\der_{\udl{5}}$ 
because $\abs{\brkt{\der_{\udl{4}}+i\der_{\udl{5}}}\cA}^2 
 = \der^m\cA^\dagger\der_m\cA$. 
Since 
\be
 \der_{\udl{4}}+i\der_{\udl{5}} 
 = -\frac{i\sqrt{E_4E_5}}{e^{(2)}}\brkt{\sqrt{\frac{E_5}{E_4}}\der_4
 -\sqrt{\frac{E_4}{E_5}}\der_5},  
\ee
we define $\der_E$ as
\be
 \der_E \equiv \frac{1}{S_E}\der_4-S_E\der_5.  \label{def:der_E}
\ee
Then, its lowest component is 
\be
 \der_E| = \frac{ie^{(2)}}{\sqrt{E_4E_5}}\brkt{\der_{\udl{4}}+i\der_{\udl{5}}}.  
 \label{lc:der_E}
\ee
Here and hereafter, the symbol~$|$ denotes the lowest component of a superfield. 
This promoted derivative~$\der_E$ is certainly reduced to the global SUSY one~$\der$ 
if we replace $S_E$ with its background value~$s$. 

From the counting of the Weyl and chiral weights, (\ref{cL_hyp:2}) should be 
modified as
\bea
 \cL_{\rm H} \eql -\int\dr^4\tht\;2V_E^{1/2}U_E(S_E,\bar{S}_E)
 \brkt{H_{\rm odd}^\dagger\tl{d}e^V H_{\rm odd}
 +H_{\rm even}^\dagger\tl{d}e^{-V}H_{\rm even}} \nonumber\\
 &&+\sbk{\int\dr^2\tht\;\brc{H_{\rm odd}^t\tl{d}\brkt{\der_E-\Sgm}H_{\rm even}
 -H_{\rm even}^t\tl{d}\brkt{\der_E+\Sgm}H_{\rm odd}}+\hc}, 
 \label{cL_hyper:3}
\eea
where $U_E(S_E,\bar{S}_E)$ is a real function. 
From (\ref{comp:S_E}), (\ref{comp:V_E}) and (\ref{comp:Phis}), 
the lowest component of the integrand in the $d^4\tht$-integral is read off as 
\bea
 \bdm{C} \defa  
 \left.V_E^{1/2}U_E(S_E,\bar{S}_E)\brkt{H_{\rm odd}^\dagger\tl{d}e^VH_{\rm odd}
 +H_{\rm even}^\dagger\tl{d}e^{-V}H_{\rm even}}\right| \nonumber\\
 \eql \sqrt{e^{(2)}}U_E\brkt{\sqrt{\frac{E_4}{E_5}},\sqrt{\frac{\bar{E}_4}{\bar{E}_5}}}
 \cdot\abs{(E_4E_5)^{1/4}}^2\brkt{\cA_{\rm odd}^\dagger\tl{d}\cA_{\rm odd}
 +\cA_{\rm even}^\dagger\tl{d}\cA_{\rm even}}, 
\eea
where $\cA_{\rm odd}$ and $\cA_{\rm even}$ are column vectors that consist of 
$\cA^{2A-1}_2$ and $\cA_2^{2A}$, respectively. 
Note that $\bdm{C}$ appears in front of the Ricci scalar 
when the $d^4\tht$-integral is promoted to the D-term action formula~\cite{Kugo:1982cu}. 
From the component expression of 6D SUGRA~\cite{Bergshoeff:1985mz}, 
on the other hand, the coefficient of the Ricci scalar 
should be $e^{(2)}\brkt{\cA_{\rm odd}^\dagger\tl{d}\cA_{\rm odd}
+\cA_{\rm even}^\dagger\tl{d}\cA_{\rm even}}$.\footnote{
Note that $\det(e_M^{\;\;\udl{N}})=e^{(2)}$ under our assumption. 
} 
Thus the function~$U_E|$ is determined as
\bea
 U_E^2| \eql \frac{e^{(2)}}{\abs{E_4E_5}} 
 = -\frac{i}{2\abs{E_4E_5}}\brkt{\bar{E}_4E_5-E_4\bar{E}_5} \nonumber\\
 \eql -\frac{i}{2}\brkt{\sqrt{\frac{\bar{E}_4E_5}{E_4\bar{E}_5}}
 -\sqrt{\frac{\bar{E}_5E_4}{E_5\bar{E}_4}}} 
 = \left.\Im\frac{\bar{S}_E}{S_E}\right|.  \label{lc:U_E}
\eea
Therefore, we obtain 
\be
 U_E(S_E,\bar{S}_E) = \brkt{\Im\frac{\bar{S}_E}{S_E}}^{1/2}. 
 \label{def:U_E}
\ee
In fact, substituting (\ref{def:U_E}) into (\ref{cL_hyper:3}) and eliminating the F-terms 
of $H_{\rm odd,even}$, we obtain the correct kinetic terms. 
\bea
 \cL_{\rm H} \eql 2e^{(2)}\brc{\der^M\cA_{\rm odd}^\dagger\tl{d}
 \der_M\cA_{\rm odd}+\der^M\cA_{\rm even}^\dagger\tl{d}\der_M\cA_{\rm even}}
 +\cdots. 
\eea

Correspondingly to the promotion:~$\der\to\der_E$, 
(\ref{compid:vector:1}) is also modified as 
\bea
 V \eql -(\tht\sgm^\mu\bar{\tht})A_\mu+\cO(\tht^3), \nonumber\\
 \Sgm \eql \brkt{\sqrt{\frac{E_5}{E_4}}A_4-\sqrt{\frac{E_4}{E_5}}A_5}+\cO(\tht) 
 \nonumber\\
 \eql \frac{ie^{(2)}}{\sqrt{E_4E_5}}\brkt{A_{\udl{4}}+iA_{\udl{5}}}+\cO(\tht). 
\eea

\subsection{Vector-tensor sector}
Next we consider the vector-tensor sector. 
The definition of the tensor (field-strength) superfield~$\Phi_T$ 
is unchanged from (\ref{def:PhiW_T}), 
\be
 \Phi_T \equiv -2iD^\alp\bar{D}^2Y_\alp+2i\bar{D}_{\dalp}D^2\bar{Y}^{\dalp},  
 \label{def:Phi_T:local}
\ee
while that of $\cW_{T\alp}$ is now modified from (\ref{def:cW_T:global}) as 
\be 
 \cW_{T\alp} \equiv \bar{D}^2\brkt{\frac{1}{S_E}D_\alp X_4
 +S_E D_\alp X_5+4S_E\cO_E Y_\alp}, 
\ee
where $X_4$ and $X_5$ are real superfields, and 
\be
 \cO_E \equiv \frac{1}{S_E^2}\der_4+\der_5. 
\ee
The constraint~(\ref{rel:XYZ:2}) is promoted to the SUGRA version: 
\be
 \bar{D}^2\brkt{\frac{1}{S_E}D_\alp X_4
 -S_E D_\alp X_5+4\der_E Y_\alp} = 0.  \label{rel:XYZ:local}
\ee
Under this constraint, $\cW_{T\alp}$ can be rewritten as 
\bea
 \cW_{T\alp} \eql \frac{1}{S_E}\cW_{4\alp}
 +\frac{8}{S_E}\der_4\bar{D}^2Y_\alp \nonumber\\
 \eql S_E\cW_{5\alp}+8S_E\der_5\bar{D}^2Y_\alp, \label{def:cW_T:3}
\eea
which is the SUGRA version of (\ref{def:cW_T:2}). 
The field strength superfields~$\cW_{4\alp}$ and $\cW_{5\alp}$ 
are defined as (\ref{def:cW_45}). 
The superfields~$\Phi_T$, $\cW_{T\alp}$ and the constraint~(\ref{rel:XYZ:local}) 
are invariant under the gauge transformation:
\bea
 \dlt X_4 \eql \der_4 V_G-\Re(S_E\Sgm_G), \;\;\;\;\;
 \dlt X_5 = \der_5 V_G+\Re\brkt{\frac{\Sgm_G}{S_E}}, \nonumber\\
 \dlt Y_\alp \eql -\frac{1}{4}D_\alp V_G,  \label{gauge_trf:tensor}
\eea 
where the transformation parameters~$V_G$ and $\Sgm_G$ are a real and a chiral superfields 
that form a 6D vector multiplet. 
From the expressions in (\ref{def:Phi_T:local}) and (\ref{def:cW_T:3}), 
we can show that 
\be
 D^\alp\brkt{U_E^2\cW_{T\alp}} = -2\bar{\der}_E\Phi_T
 +\frac{i\bar{D}_{\dalp}\bar{S}_E}{\bar{S}_E}\bar{\cW}_T^{\dalp}, 
 \label{SGconstraint1}
\ee
which is the SUGRA extension of the first constraint in (\ref{constraints}). 
From the gauge invariance of the action, the second constraint in (\ref{constraints}) 
should be modified as
\be
 \bar{D}^2D_\alp(V_E\Phi_T) = -4\brc{\der_E\cW_{T\alp}
 -(\cO_ES_E)\cW_{T\alp}}.  \label{SGconstraint2}
\ee
(See Sec.~\ref{gauge_inv}.) 
The bosonic components of $X_4$, $X_5$ and $Y_\alp$ are given by
\bea
 X_4 \eql \frac{1}{4}(\tht\sgm^\mu\bar{\tht})B_{\mu 4}+\cdots, \;\;\;\;\;
 X_5 = \frac{1}{4}(\tht\sgm^\mu\bar{\tht})B_{\mu 5}+\cdots, \nonumber\\
 Y_\alp \eql \frac{1}{16}\tht_\alp\bar{\tht}^2\brkt{B_{45}+\frac{i}{2}\sgm} 
 +\frac{i}{16}(\sgm^{\mu\nu}\tht)_\alp\bar{\tht}^2 B_{\mu\nu}+\cdots, 
\eea
where $B_{MN}$ is an unconstrained tensor field. 
\ignore{
From these expressions, we obtain 
\bea
 \Phi_T \eql \sgm+(\tht\sgm^\mu\bar{\tht})\ep_{\mu\nu\rho\lmd}
 \der^\nu B^{\rho\lmd}-\frac{1}{4}\tht^2\bar{\tht}^2\Box_4\sgm+\cdots, \nonumber\\
 \cW_{T\alp} \eql \tht_\alp\bar{\der}_E|\sgm
 +(\sgm^{\mu\nu}\tht)_\alp\brc{\bar{\der}_E|B_{\mu\nu}
 +\der_\mu C_\nu-\der_\nu C_\mu}+\cdots, 
\eea
where the symbol~$|$ denotes the lowest component in the superfields, and 
$C_\mu\equiv -i\brkt{S_E^{-1}|B_{\mu 4}+S_E|B_{\mu 5}}$. 
}

As explained in Appendix~\ref{comp_of_constraint}, 
the constraint~(\ref{rel:XYZ:local}) can be satisfied 
for arbitrary unconstrained superfields~$Y_\alp$ and $X_4$ 
by adjusting $S_E$ and $X_5$. 
This indicates that the latter two superfields are not independent. 
In fact, we can express the action without $X_5$ by adopting 
the first equation in (\ref{def:cW_T:3}) as the definition of $\cW_{T\alp}$. 
This reflects the fact that $X_5$ can be gauged away by (\ref{gauge_trf:tensor}). 
Of course, we can choose $Y_\alp$ and $X_5$ as independent superfields. 

Now we promote $\cL_{\rm VT}$ in (\ref{cL:global:2}) to SUGRA 
by replacing $\der$ with $\der_E$ 
and inserting $V_E$ to match the Weyl weight of the integrand to 2, and obtain 
\bea
 \cL_{\rm VT} \eql \int\dr^4\tht\;f_{IJ}\left[
 \brc{-2\Sgm^ID^\alp V^J\cW_{T\alp}
 +\frac{1}{2}\brkt{\der_E V^I D^\alp V^J
 -\der_E D^\alp V^IV^J}\cW_{T\alp}+\hc} \right.\nonumber\\
 &&\hspace{20mm}
 +\Phi_TV_E\brkt{D^\alp V^I\cW_\alp^I+\bar{D}_{\dalp}V^I\bar{\cW}^{J\dalp}
 +V^ID^\alp\cW_\alp^J} \nonumber\\
 &&\hspace{20mm}\left.
 +\frac{\Phi_T}{U_E^2}\brc{4(\bar{\der}_EV^I-\bar{\Sgm}^I)(\der_E V^J-\Sgm^J)
 -2\bar{\der}_EV^I\der_E V^J}\right]. 
 \label{cL_VT:local}
\eea
The factor~$U_E^{-2}$ is necessary in order to obtain the correct component expression 
of the Lagrangian. 
Note that the third line in (\ref{cL_VT:local}) provides the extra-dimensional 
components of the kinetic terms for the 6D vector fields. 
The lowest component of $U_E^{-2}$ 
cancels the unwanted factor in (\ref{lc:der_E}). 

In order for the Lagrangian to be gauge-invariant, we need to add 
the following terms to (\ref{cL_VT:local}). (See Sec.~\ref{gauge_inv}.) 
\bea
 \cL_{\Sgm^2}^{\rm (SG)} \eql \int\dr^4\tht\;2f_{IJ}\frac{\Phi_T}{U_E^2}
 \brkt{\frac{S_E}{\bar{S}_E}\Sgm^I\Sgm^J
 +\frac{\bar{S}_E}{S_E}\bar{\Sgm}^I\bar{\Sgm}^J}.  \label{cL_add}
\eea
Note that this vanishes if $V_E$ and $S_E$ are replaced with 
their background values.

\subsection{6D SUGRA action} \label{expr:action}
In summary, the 6D SUGRA action is expressed as
\bea
 S^{\rm (SG)} \eql \int\dr^6x\;\brkt{\cL_{\rm H}^{\rm (SG)}
 +\cL_{\rm VT}^{\rm (SG)}}, \nonumber\\
 \cL^{\rm (SG)}_{\rm H} \eql -\int\dr^4\tht\;2V_E^{1/2}U_E(S_E,\bar{S}_E)
 \brkt{H_{\rm odd}^\dagger\tl{d}e^V H_{\rm odd}
 +H_{\rm even}^\dagger\tl{d}e^{-V}H_{\rm even}} \nonumber\\
 &&+\sbk{\int\dr^2\tht\;\brc{H_{\rm odd}^t\tl{d}\brkt{\der_E-\Sgm}H_{\rm even}
 -H_{\rm even}^t\tl{d}\brkt{\der_E+\Sgm}H_{\rm odd}}+\hc}, \nonumber\\
 \cL^{\rm (SG)}_{\rm VT} \eql \int\dr^4\tht\;f_{IJ}\left[
 \brc{-2\Sgm^ID^\alp V^J\cW_{T\alp}
 +\frac{1}{2}\brkt{\der_E V^I D^\alp V^J
 -\der_E D^\alp V^IV^J}\cW_{T\alp}+\hc} \right.\nonumber\\
 &&\hspace{20mm}
 +\Phi_TV_E\brkt{D^\alp V^I\cW_\alp^J+\bar{D}_{\dalp}V^I\bar{\cW}^{J\dalp}
 +V^ID^\alp\cW_\alp^J} \nonumber\\
 &&\hspace{20mm}
 +\frac{\Phi_T}{U_E^2}\left\{4(\bar{\der}_EV^I-\bar{\Sgm}^I)(\der_E V^J-\Sgm^J)
 -2\bar{\der}_EV^I\der_E V^J \right. \nonumber\\
 &&\hspace{30mm}\left.\left.
 +\frac{2S_E}{\bar{S}_E}\Sgm^I\Sgm^J
 +\frac{2\bar{S}_E}{S_E}\bar{\Sgm}^I\bar{\Sgm}^J\right\}\right]. 
 \label{cL:SUGRA}
\eea
This certainly reproduces the global SUSY action in the previous section 
when $V_E=1$ and $S_E=s$. 

\ignore{
Under an exchange~$x^4\exch x^5$, each superfield transforms as
\bea
 &&S_E \exch -S_E^{-1}, \;\;\;\;\;
 V_E^{1/2} \exch iV_E^{1/2}, \;\;\;\;\;
 V^I \exch V^I, \;\;\;\;\;
 \Sgm^I \exch \Sgm^I, \nonumber\\
 &&H_{\rm odd} \exch H_{\rm odd}, \;\;\;\;\;
 H_{\rm even} \exch H_{\rm even}, \;\;\;\;\;
 X_4 \exch X_5, \;\;\;\;\;
 Y_\alp \exch -Y_\alp. \label{flip:sf}
\eea
This leads to~\footnote{
We have chosen branches of square roots in the definitions 
of $S_E$, $V_E$ and $U_E$ so that (\ref{flip:sf}) 
(\ref{flip:UE}) are realized. 
} 
\be
 U_E \exch -i U_E, \;\;\;\;\;
 \der_E \exch \der_E.  \label{flip:UE}
\ee
Hence, noting that 
\bea
 \int\dr^4\tht\;\frac{\Phi_T}{U_E^2}\frac{S_E}{\bar{S}_E}\Sgm^I\Sgm^J 
 &\exch & \int\dr^4\tht\;\frac{\Phi_T}{U_E^2}\frac{\bar{S}_E}{S_E}\Sgm^I\Sgm^J \nonumber\\
 \eql \int\dr^4\tht\;\frac{\Phi_T}{U_E^2}\brkt{2iU_E^2+\frac{S_E}{\bar{S}_E}}\Sgm^I\Sgm^J 
 \nonumber\\
 \eql \int\dr^4\tht\;\frac{\Phi_T}{U_E^2}\frac{S_E}{\bar{S}_E}\Sgm^I\Sgm^J, 
\eea 
we find that $\cL^{\rm (SG)}\exch \cL^{\rm (SG)}$. 
Thus the action~$S^{\rm (SG)}$ is invariant under this exchange 
because $d^6x\exch d^6x$.} 

\ignore{
We also note that $\cL_{\rm LM}^{\rm (SG)}$ can be rewritten as
\bea
 \cL_{\rm LM}^{\rm (SG)} \eql \int\dr^4\tht\;
 \brc{V_E\Phi_T\tl{\Phi}_T-8i\der_E\tl{Y}^\alp\cW_{T\alp}
 +8i\bar{\der}_E\bar{\tl{Y}}_{\dalp}\bar{\cW}_T^{\dalp}},  \label{cL_LM:local:2}
\eea
where $\tl{\Phi}_T\equiv -2iD^\alp\bar{D}^2\tl{Y}_\alp
+2i\bar{D}_{\dalp}D^2\bar{\tl{Y}}^{\dalp}$ 
and used the first formula in (\ref{fml:AB}). 
If we identify $\tl{Y}_\alp$ as a superfield coming from another 
6D tensor multiplet, we can understand (\ref{cL_LM:local:2}) as 
the $\cN=1$ superfield description of (3.53) of Ref.~\cite{Linch:2012zh}
}

\ignore{
We comment that $V_E$ can be integrated out, 
just like in 5D SUGRA
The equation of motion for $V_E$ is
\bea
 &&-V_E^{-1/2}U_E\Omg_{\rm H}
 +f_{IJ}\Phi_T\brkt{D^\alp V^I\cW_\alp^I+\bar{D}_{\dalp}V^I\bar{\cW}^{J\dalp}
 +V^ID^\alp\cW^J_\alp} \nonumber\\
 &&+\Phi_T\brkt{-2iD^\alp\bar{D}^2\tl{Y}_\alp+\hc} = 0, 
\eea
where 
\be
 \Omg_{\rm H} \equiv \Phi_{\rm odd}^\dagger\tl{d}e^V\Phi_{\rm odd}
 +\Phi_{\rm even}^\dagger\tl{d}e^{-V}\Phi_{\rm even}. 
\ee
After eliminating $V_E$ from the Lagrangian by this, we obtain 
\bea
 \cL^{\rm (SG)} \eql -\int\dr^4\tht\;\frac{U_E^2\Omg_{\rm H}^2}
 {\Phi_T\brc{\tl{\Phi}_T+f_{IJ}
 \brkt{D^\alp V^I\cW_\alp^I+\bar{D}_{\dalp}V^I\bar{\cW}^{J\dalp}+V^ID^\alp\cW^J_\alp}}} 
 \nonumber\\
 &&+\sbk{\int\dr^2\tht\;\brc{\Phi_{\rm odd}^t\tl{d}\brkt{\der_E-\Sgm}\Phi_{\rm even}
 -\Phi_{\rm even}^t\tl{d}\brkt{\der_E+\Sgm}\Phi_{\rm odd}}+\hc} \nonumber\\
 &&-\int\dr^4\tht\;f_{IJ}
 \brc{2\Sgm^ID^\alp V^J+8i\der_E\tl{Y}^\alp
 -\frac{1}{2}\brkt{\der_E V^I D^\alp V^J
 -\der_E D^\alp V^IV^J}}\cW_{T\alp}+\hc \nonumber\\
 &&+\int\dr^4\tht\;
 \frac{f_{IJ}\Phi_T}{U_E^2}\left\{4(\bar{\der}_EV^I-\bar{\Sgm}^I)(\der_E V^J-\Sgm^J)
 -2\bar{\der}_EV^I\der_E V^J \right.\nonumber\\
 &&\hspace{30mm}\left.
 +\frac{2S_E}{\bar{S}_E}\Sgm^I\Sgm^J
 +\frac{2\bar{S}_E}{S_E}\bar{\Sgm}^I\bar{\Sgm}^J\right\}, 
\eea
where $\tl{\Phi}_T\equiv -2iD^\alp\bar{D}^2\tl{Y}_\alp+\hc$, 
and we have used (\ref{fml:AB}). 
}

Here we comment on 
the constraints~(\ref{rel:XYZ:local}) and (\ref{SGconstraint2}). 
They can be released by introducing the following terms. 
\bea
 \cL_{\rm LM}^{\rm (SG)} \eql \int\dr^4\tht\;i\tl{Z}^\alp\bar{D}^2
 \brkt{\frac{1}{S_E}D_\alp X_4-S_E D_\alp X_5+4\der_E Y_\alp} \nonumber\\
 &&+\int\dr^4\tht\;2i\tl{Y}^\alp\sbk{\bar{D}^2D_\alp(V_E\Phi_T)
 +4\brc{\der_E\cW_{T\alp}-(\cO_ES_E)\cW_{T\alp}}}+\hc, 
 \label{cL_LM:1}
\eea
where the Lagrange multipliers~$\tl{Z}^\alp$ and $\tl{Y}^\alp$ 
are unconstrained superfields.\footnote{
If we identify $\tl{Y}_\alp$ as a superfield coming from another 6D tensor multiplet, 
we can understand the second line of (\ref{cL_LM:1}) as 
the $\cN=1$ superfield description of (3.53) of Ref.~\cite{Linch:2012zh}, 
which is described in the projective superspace.  
} 
These terms can be rewritten as
\bea
 \cL_{\rm LM}^{\rm (SG)} \eql \int\dr^4\tht\;i\brc{D^\alp\bar{D}^2(S_E\tl{Z}_\alp)
 -\bar{D}_{\dalp}D^2(\bar{S}_E\bar{\tl{Z}}^{\dalp})}X_5 \nonumber\\
 &&+\int\dr^4\tht\;\brc{i(S_E\tl{Z}^\alp)
 \brkt{\frac{1}{2S_E^2}\cW_{4\alp}+\frac{4}{S_E}\der_E\bar{D}^2Y_\alp}+\hc}
 \nonumber\\
 &&+\int\dr^4\tht\;\brc{V_E\Phi_T\tl{\Phi}_T-8i\der_E\tl{Y}^\alp\cW_{T\alp}
 +8i\bar{\der}_E\bar{\tl{Y}}_{\dalp}\bar{\cW}_T^{\dalp}}, 
 \label{cL_LM:2}
\eea
where $\tl{\Phi}_T\equiv -2iD^\alp\bar{D}^2\tl{Y}_\alp
+2i\bar{D}_{\dalp}D^2\bar{\tl{Y}}^{\dalp}$. 
We have dropped total derivatives. 
If we adopt the first equation in (\ref{def:cW_T:3}) as 
the definition of $\cW_{T\alp}$, a real superfield~$X_5$ only appears 
in the first line of (\ref{cL_LM:2}) and thus is regarded as 
a Lagrange multiplier. 
Then its equation of motion provides 
\be
 D^\alp\bar{D}^2(S_E\tl{Z}_\alp) = \bar{D}_{\dalp}D^2(\bar{S}_E\bar{\tl{Z}}^{\dalp}), 
\ee
which is understood as the Bianchi identity. 
Thus, this can be solved as
\be
 S_E\tl{Z}_\alp = \frac{1}{2}D_\alp V_Z, 
\ee
where $V_Z$ is a real superfield. 
Therefore, (\ref{cL_LM:2}) is rewritten as
\bea
 \cL_{\rm LM}^{\rm (SG)} \eql \left[\int\dr^2\tht\;\left\{\frac{i}{4S_E^2}\cW_Z\cW_4
 +\frac{2i}{S_E}\brkt{\cW_Z^\alp\der_E\bar{D}^2Y_\alp
 +\cW_4^\alp\der_E\bar{D}^2\tl{Y}_\alp} \right.\right.\nonumber\\
 &&\hspace{15mm}\left.\left.
 +\frac{16i}{S_E}\der_E\bar{D}^2\tl{Y}^\alp\der_4\bar{D}^2Y_\alp\right\}+\hc\right]
 +\int\dr^4\tht\;V_E\Phi_T\tl{\Phi}_T, 
\eea
where $\cW_{Z\alp}\equiv -\frac{1}{4}\bar{D}^2D_\alp V_Z$. 
Note that all the superfields are now unconstrained in this expression. 
Needless to say, we can choose $X_4$ instead of $X_5$ 
as the Lagrange multiplier, and adopt the second equation in (\ref{def:cW_T:3}) 
as the definition of $\cW_{T\alp}$.

\section{Consistency checks}  \label{consistency_checks}
In this section, we show that our result~(\ref{cL:SUGRA}) is 
gauge-invariant, and is reduced to the known superfield expression 
of 5D SUGRA after the dimensional reduction. 

\subsection{Gauge invariance} \label{gauge_inv}
The (super)gauge transformation is given by 
\bea
 V_E \toa V_E, \;\;\;\;\;
 S_E \to S_E, \nonumber\\
 H_{\rm odd} \toa e^{-\Lmd}H_{\rm odd}, \;\;\;\;\;
 H_{\rm even} \to e^\Lmd H_{\rm even}, \nonumber\\
 V^I \toa V^I+ \Lmd^I+\bar{\Lmd}^I, \;\;\;\;\;
 \Sgm^I \to \Sgm^I+\der_E\Lmd^I, \nonumber\\
 Y_\alp \toa Y_\alp, \;\;\;\;\;
 X_4 \to X_4, \;\;\;\;\;
 X_5 \to X_5.  \label{gauge_trf:local}
\eea 
Under this transformation, $\cL_{\rm H}^{(\rm SG)}$ is manifestly invariant, 
while the invariance of the remaining part~$\cL_{\rm VT}^{(\rm SG)}$ 
is quite nontrivial because it is invariant only up to total derivatives. 
In the following, we neglect total derivative terms. 
Note that the following formulae hold. 
\bea
 (\der_E A)B \eql -A\der_E B+(\cO_E S_E)AB, \nonumber\\
 D^\alp\der_E A \eql \der_E D^\alp A-(D^\alp S_E)\cO_E A.  \label{fml:AB}
\eea
The variation of $\cL_{\rm VT}^{(\rm SG)}$ is 
\bea
 \dlt\cL^{(\rm SG)}_{\rm VT} \eql \int\dr^4\tht\;f_{IJ}\left[\left\{
 -2\der_E\Lmd^I D^\alp V^J\cW_{T\alp}-2\Sgm^I D^\alp\Lmd^J\cW_{T\alp} 
 \right.\right. \nonumber\\
 &&\hspace{20mm}
 +\frac{1}{2}\left\{\der_E\brkt{\Lmd^I+\bar{\Lmd}^I}D^\alp V^J
 +\der_E V^ID^\alp\Lmd^J-\der_E D^\alp\Lmd^I V^J \right.\nonumber\\
 &&\hspace{25mm}\left.\left.
 -\der_E D^\alp V^I\brkt{\Lmd^J+\bar{\Lmd}^J}\right\}\cW_{T\alp}+\hc\right\} \nonumber\\
 &&\hspace{15mm}
 +\Phi_TV_E\brc{D^\alp\Lmd^I\cW_\alp^J+\bar{D}_{\dalp}\bar{\Lmd}^I\bar{\cW}^{J\dalp}
 +\brkt{\Lmd^I+\bar{\Lmd}^I}D^\alp\cW_\alp^J} \nonumber\\
 &&\hspace{15mm}
 +\frac{\Phi_T}{U_E^2}\left\{
 4\bar{\der}_E\Lmd^I\brkt{\der_E V^J-\Sgm^J}
 +4\brkt{\bar{\der}_E V^I-\bar{\Sgm}^I}\der_E\bar{\Lmd}^J \right. \nonumber\\
 &&\hspace{25mm}
 -2\bar{\der}_E\brkt{\Lmd^I+\bar{\Lmd}^I}\der_E V^J
 -2\bar{\der}_E V^I\der_E\brkt{\Lmd^J+\bar{\Lmd}^J}  \nonumber\\
 &&\hspace{25mm}\left.\left.
 +\frac{4S_E}{\bar{S}_E}\der_E\Lmd^I\Sgm^J
 +\frac{4\bar{S}_E}{S_E}\bar{\der}_E\bar{\Lmd}^I\bar{\Sgm}^J \right\}\right] \nonumber\\
 \eql \int\dr^4\tht\;f_{IJ}\left[
 \frac{1}{2}\left\{
 \der_E\brkt{-3\Lmd^I+\bar{\Lmd}^I}D^\alp V^J
 +D^\alp\Lmd^I\der_E V^J \right.\right. \nonumber\\
 &&\hspace{25mm}\left.
 -\der_E D^\alp\Lmd^I V^J
 -\brkt{\Lmd^I+\bar{\Lmd}^I}\der_E D^\alp V^J\right\}\cW_{T\alp} 
 \nonumber\\
 &&\hspace{20mm}
 +\Phi_TV_E\brkt{D^\alp\Lmd^I\cW_\alp^J+\Lmd^ID^\alp\cW_\alp^J}  \nonumber\\
 &&\hspace{20mm}\left.
 +\frac{2\Phi_T}{U_E^2}\bar{\der}_E\brkt{\Lmd^I-\bar{\Lmd}^I}\der_E V^J
 +\hc \right].  \label{dltcL_VT:1}
\eea
At the second equality, we have used the following equation:  
\bea
 \int\dr^4\tht\;\frac{\Phi_T}{U_E^2}\frac{\bar{S}_E}{S_E}\der_E\Lmd^I\Sgm^J 
 \eql \int\dr^4\tht\;\frac{\Phi_T}{U_E^2}\frac{\bar{S}_E}{S_E}
 \brkt{\frac{1}{S_E}\der_4-S_E\der_5}\Lmd^I\Sgm^J \nonumber\\
 \eql \int\dr^4\tht\;\frac{\Phi_T}{U_E^2}
 \brc{\brkt{2iU_E^2+\frac{S_E}{\bar{S}_E}}\frac{1}{S_E}\der_4\Lmd^I
 -\bar{S}_E\der_5\Lmd^I}\Sgm^J \nonumber\\
 \eql \int\dr^4\tht\;\Phi_T\bar{\der}_E\Lmd^I\Sgm^J. 
\eea
The last equality holds because of the property of $\Phi_T$ as a linear superfield. 
By means of (\ref{fml:AB}), we can show that
\bea
 &&\frac{1}{2}\brc{\der_E\brkt{-3\Lmd^I+\bar{\Lmd}^I}D^\alp V^J
 +D^\alp\Lmd^I\der_E V^J
 -\der_E D^\alp\Lmd^I V^J
 -\brkt{\Lmd^I+\bar{\Lmd}^I}\der_E D^\alp V^J} \nonumber\\
 \eql \frac{1}{2}\left\{
 \der_E D^\alp\brkt{\bar{\Lmd}^I-\Lmd^I}V^J
 +\der_E\brkt{\bar{\Lmd}^I-\Lmd^I}D^\alp V^J \right. \nonumber\\
 &&\hspace{5mm}\left.
 -D^\alp\brkt{\bar{\Lmd}^I-\Lmd^I}\der_E V^J
 -\brkt{\bar{\Lmd}^I-\Lmd^I}\der_E D^\alp V^J\right\} 
 -\der_E\Lmd^I D^\alp V^J-\Lmd^I\der_E D^\alp V^J \nonumber\\
 \eql \frac{1}{2}D^\alp\brc{\der_E\brkt{\bar{\Lmd}^I-\Lmd^I}V^J
 -\brkt{\bar{\Lmd}^I-\Lmd^I}\der_E V^J} \nonumber\\
 &&+\frac{D^\alp S_E}{2}\brc{\cO_E\brkt{\bar{\Lmd}^I-\Lmd^I}V^J
 -\brkt{\bar{\Lmd}^I-\Lmd^I}\cO_E V^J}
 -\der_E\brkt{\Lmd^I D^\alp V^J}. 
\eea
Thus (\ref{dltcL_VT:1}) is rewritten as 
\bea
 \dlt\cL_{\rm VT}^{(\rm SG)} \eql \int\dr^4\tht\;f_{IJ}\left[
 \frac{1}{2}D^\alp\brc{\der_E\brkt{\bar{\Lmd}^I-\Lmd^I}V^J
 -\brkt{\bar{\Lmd}^I-\Lmd^I}\der_E V^J}\cW_{T\alp} \right. \nonumber\\
 &&\hspace{20mm}
 +\frac{D^\alp S_E}{2}\brc{\cO_E\brkt{\bar{\Lmd}^I-\Lmd^I}V^J
 -\brkt{\bar{\Lmd}^I-\Lmd^I}\cO_E V^J}\cW_{T\alp} \nonumber\\
 &&\hspace{20mm}
 -\der_E\brkt{\Lmd^I D^\alp V^J}\cW_{T\alp}
 +V_E\Phi_T D^\alp\brkt{\Lmd^I\cW^J_\alp} \nonumber\\
 &&\hspace{20mm}\left.
 +\frac{2\Phi_T}{U_E^2}\bar{\der}_E\brkt{\Lmd^I-\bar{\Lmd}^I}\der_E V^J+\hc\right] 
 \nonumber\\
 \eql \int\dr^4\tht\;f_{IJ}\left[
 \frac{1}{2}\brc{\der_E\brkt{\Lmd^I-\bar{\Lmd}^I}V^J
 -\brkt{\Lmd^I-\bar{\Lmd}^I}\der_E V^J}D^\alp\cW_{T\alp} \right.\nonumber\\
 &&\hspace{20mm}
 -\frac{D^\alp S_E}{2}\brc{\cO_E\brkt{\Lmd^I-\bar{\Lmd}^I}V^J
 -\brkt{\Lmd^I-\bar{\Lmd}^I}\cO_E^\alp V^J}\cW_{T\alp} \nonumber\\
 &&\hspace{20mm}\left.
 +\frac{2\Phi_T}{U_E^2}\bar{\der}_E\brkt{\Lmd^I-\bar{\Lmd}^I}\der_E V^J+\hc \right]. 
 \label{dltcL_VT:2}
\eea
At the second equality, we have used that
\bea
 &&-\der_E\brkt{\Lmd^I D^\alp V^J}\cW_{T\alp}
 +V_E\Phi_T D^\alp\brkt{\Lmd^I\cW^J_\alp} \nonumber\\
 \eql \Lmd^I D^\alp V^J\brc{\der_E\cW_{T\alp}-(\cO_E S_E)\cW_{T\alp}}
 -D^\alp\brkt{V_E\Phi_T}\Lmd^I\cW^J_\alp \nonumber\\
 \eql \frac{1}{4}\Lmd^ID^\alp V^J\brc{
 \bar{D}^2D_\alp(V_E\Phi_T)+4\brkt{\der_E\cW_{T\alp}-(\cO_E S_E)\cW_{T\alp}}}
 = 0,  
\eea
where (\ref{SGconstraint2}) is used at the last step. 
\ignore{
The last line in (\ref{dltcL_VT:2}) can be canceled with 
the variation of $\cL_{\rm LM}^{(\rm SG)}$ in (\ref{cL_add}) 
if we assume the gauge transformation of $\tl{Y}^\alp$ to be
\be
 \dlt\tl{Y}^\alp = \frac{if_{IJ}}{8}\Lmd^ID^\alp V^J. 
\ee
}

Using (\ref{def:cW_T:3}), we find that 
\bea
 &&\frac{1}{2}\der_E\brkt{\Lmd^I-\bar{\Lmd}^I}V^J D^\alp\cW_{T\alp}+\hc \nonumber\\
 \eql \frac{1}{2S_E}\der_4\brkt{\Lmd^I-\bar{\Lmd}^I}V^J 
 D^\alp\brc{S_E\brkt{\cW_{5\alp}+8\der_5\bar{D}^2Y_\alp}} \nonumber\\
 &&-\frac{S_E}{2}\der_5\brkt{\Lmd^I-\bar{\Lmd}^I}V^J 
 D^\alp\brc{\frac{1}{S_E}\brkt{\cW_{4\alp}+8\der_4\bar{D}^2Y_\alp}}+\hc 
 \nonumber\\
 \eql \frac{D^\alp S_E}{2}\cO_E\brkt{\Lmd^I-\bar{\Lmd}^I}V^J\cW_{T\alp} \nonumber\\
 &&+\frac{1}{2}\der_4\brkt{\Lmd^I-\bar{\Lmd}^I}V^J
 D^\alp\brkt{\cW_{5\alp}+8\der_5\bar{D}^2Y_\alp} \nonumber\\
 &&-\frac{1}{2}\der_5\brkt{\Lmd^I-\bar{\Lmd}^I}V^J 
 D^\alp\brkt{\cW_{4\alp}+8\der_4\bar{D}^2Y_\alp}+\hc \nonumber\\
 \eql \brc{\frac{D^\alp S_E}{2}\cO_E\brkt{\Lmd^I-\bar{\Lmd}^I}V^J\cW_{T\alp}+\hc} 
 \nonumber\\
 &&+2i\der_4\brkt{\Lmd^I-\bar{\Lmd}^I}V^J\der_5\Phi_T
 +2i\der_5\brkt{\Lmd^I-\bar{\Lmd}^I}V^J\der_4\Phi_T. 
\eea
Similarly, we obtain
\bea
 -\frac{1}{2}\brkt{\Lmd^I-\bar{\Lmd}^I}\der_E V^J D^\alp\cW_{T\alp}+\hc 
 \eql \brc{-\frac{D^\alp S_E}{2}\brkt{\Lmd^I-\bar{\Lmd}^I}\cO_E V^J\cW_{T\alp}+\hc} 
 \nonumber\\
 &&-2i\brkt{\Lmd^I-\bar{\Lmd}^I}
 \brc{\der_4 V\der_5\Phi_T-\der_5 V\der_4\Phi_T}. 
\eea
Furthermore, we can see that 
\bea
 &&\frac{2\Phi_T}{U_E^2}\Phi_T\bar{\der}_E\brkt{\Lmd^I-\bar{\Lmd}^I}\der_E V^J+\hc
 \nonumber\\
 \eql 4i\Phi_T\brc{\brkt{\Lmd^I-\bar{\Lmd}^I}\der_5 V^J
 -\der_5\brkt{\Lmd^I-\bar{\Lmd}^I}\der_4 V^J}. 
\eea
By means of these equations, we find that
\bea
 \dlt\cL_{\rm VT}^{\rm (SG)} \eql \int\dr^4\tht\;f_{IJ}\left[
 2iV^I\brc{\der_4\brkt{\Lmd^J-\bar{\Lmd}^J}\der_5\Phi_T
 -\der_5\brkt{\Lmd^J-\bar{\Lmd}^J}\der_4\Phi_T} \right.\nonumber\\
 &&\hspace{20mm}
 -2i\brkt{\Lmd^I-\bar{\Lmd}^I}
 \brkt{\der_4 V^J\der_5\Phi_T-\der_5 V^J\der_4\Phi_T} \nonumber\\
 &&\hspace{20mm}\left.
 +4i\Phi_T\brc{\der_4\brkt{\Lmd^I-\bar{\Lmd}^I}\der_5 V^J
 -\der_5\brkt{\Lmd^I-\bar{\Lmd}^I}\der_4 V^J}\right]  \nonumber\\
 \eql \int\dr^4\tht\;f_{IJ}\left[
 -2i\Phi_T\brc{\der_5 V^I\der_4\brkt{\Lmd^J-\bar{\Lmd}^J}
 -\der_4 V^I\der_5\brkt{\Lmd^J-\bar{\Lmd}^J}} \right. \nonumber\\
 &&\hspace{20mm}
 +2i\Phi_T\brc{\der_5\brkt{\Lmd^I-\bar{\Lmd}^I}\der_4 V^J
 -\der_4\brkt{\Lmd^I-\bar{\Lmd}^I}\der_5 V^J} \nonumber\\
 &&\hspace{20mm}\left. 
 +4i\Phi_T\brc{\der_4\brkt{\Lmd^I-\bar{\Lmd}^I}\der_5 V^J
 -\der_5\brkt{\Lmd^I-\bar{\Lmd}^I}\der_4 V^J}\right]  \nonumber\\
 \eql 0.  
\eea
Namely, the 6D SUGRA action~(\ref{cL:SUGRA}) is gauge-invariant.

\subsection{Dimensional reduction to 5D} \label{dim_red}
Here we show that the our result~(\ref{cL:SUGRA}) reproduces  
the known 5D SUGRA action after the dimensional reduction. 
We drop the $x^5$-dependence of the superfields in (\ref{cL:SUGRA}).\footnote{
The case that the $x^4$-dependence is dropped is essentially the same. 
}
Then the differential operators become
\be
 \der_E \to \frac{1}{S_E}\der_4, \;\;\;\;\;
 \cO_E \to \frac{1}{S_E^2}\der_4. 
\ee
Hence the hyper-sector Lagrangian~$\cL_{\rm H}^{\rm (SG)}$ in (\ref{cL:SUGRA}) becomes
\bea 
 \cL^{(\rm 5D)}_{\rm H} \eql -\int\dr^4\tht\;2V_E^{1/2}U_E
 \brc{H_{\rm odd}^\dagger\tl{d}e^V H_{\rm odd}
 +H_{\rm even}^\dagger\tl{d}e^{-V}H_{\rm even}} \nonumber\\
 &&+\sbk{\int\dr^2\tht\;\brc{H_{\rm odd}^t\tl{d}
 \brkt{\frac{1}{S_E}\der_4-\Sgm}H_{\rm even}
 -H_{\rm even}^t\tl{d}\brkt{\frac{1}{S_E}\der_4+\Sgm}H_{\rm odd}}+\hc} 
 \nonumber\\
 \eql -\int\dr^4\tht\;2\hat{V}_E\brc{
 \hat{H}_{\rm odd}^\dagger\tl{d}e^V\hat{H}_{\rm odd}
 +\hat{H}_{\rm even}^\dagger\tl{d}e^{-V}\hat{H}_{\rm even}} \nonumber\\
 &&+\sbk{\int\dr^2\tht\;\brc{\hat{H}_{\rm odd}^t\tl{d}
 \brkt{\der_4-\hat{\Sgm}}\hat{H}_{\rm even}
 -\hat{H}_{\rm even}^t\tl{d}\brkt{\der_4
 +\hat{\Sgm}}\hat{H}_{\rm odd}}+\hc},  \label{5DcL_hyp}
\eea
where 
\bea
 \hat{V}_E \defa V_E^{1/2}U_E\abs{S_E}, \;\;\;\;\;
 \hat{\Sgm}^I \equiv S_E\Sgm^I, \nonumber\\
 \hat{H}_{\rm odd} \defa S_E^{-1/2}H_{\rm odd}, \;\;\;\;\;
 \hat{H}_{\rm even} \equiv S_E^{-1/2}H_{\rm even}. \label{5Dsf}
\eea

Next we consider the vector-tensor sector Lagrangian~$\cL^{\rm (SG)}_{\rm VT}$.  
From (\ref{def:cW_T:3}), $\cW_{T\alp}$ becomes  
\be
 \cW_{T\alp} \to S_E\cW_{5\alp}. 
\ee
Then, $\cL^{\rm (SG)}_{\rm VT}$ becomes 
\bea
 \cL_{\rm VT}^{\rm (5D)} \eql 
 \int\dr^4\tht\;f_{IJ}\left[
 \brc{-2\Sgm^ID^\alp V^JS_E\cW_{5\alp}
 +\frac{1}{2}\brkt{\der_4 V^I D^\alp V^J-\der_4 D^\alp V^IV^J}\cW_{5\alp}+\hc}
 \right.\nonumber\\
 &&\hspace{20mm}
 +\Phi_TV_E\brkt{D^\alp V^I\cW_\alp^J
 +\bar{D}_{\dalp}V^I\bar{\cW}^{J\dalp}
 +V^ID^\alp\cW_\alp^J} \nonumber\\
 &&\hspace{20mm}
 +\frac{\Phi_T}{U_E^2\abs{S_E}^2}\left\{
 4\brkt{\der_4 V^I-\bar{S}_E\bar{\Sgm}^I}\brkt{\der_4-S_E\Sgm^J}
 -2\der_4 V^I\der_4 V^J \right. \nonumber\\
 &&\hspace{40mm}\left.\left.
 +2S_E^2\Sgm^I\Sgm^J+2\bar{S}_E^2\bar{\Sgm}^I\bar{\Sgm}^J\right\}\right] \nonumber\\
 \eql \int\dr^4\tht\;f_{IJ}\left[\brc{
 -2\hat{\Sgm}^ID^\alp V^J\cW_{5\alp}
 +\frac{1}{2}\brkt{\der_4 V^ID^\alp V^J-\der_4 D^\alp V^IV^J}\cW_{5\alp}+\hc} 
 \right.\nonumber\\
 &&\hspace{20mm}
 +V_E\Phi_T\brkt{D^\alp V^I\cW_\alp^J
 +\bar{D}_{\dalp}V^I\bar{\cW}^{J\dalp}
 +V^ID^\alp\cW_\alp^J} \nonumber\\
 &&\hspace{20mm}
 +\frac{2V_E\Phi_T}{\hat{V}_E^2}\left\{
 \der_4V^I\der_4V^J-2\der_4V^I\brkt{\hat{\Sgm}^J+\bar{\hat{\Sgm}}^J}
 +2\bar{\hat{\Sgm}}^I\hat{\Sgm}^J \right.\nonumber\\
 &&\hspace{40mm}\left.\left.
 +\hat{\Sgm}^I\hat{\Sgm}^J+\bar{\hat{\Sgm}}^I\bar{\hat{\Sgm}}^J\right\}\right], 
 \label{5DcL_VT:1}
\eea
where we have used (\ref{5Dsf}). 
Here, note that the constraint~(\ref{SGconstraint2}) is now
\bea
 \bar{D}^2D_\alp(V_E\Phi_T) \eql -\frac{4}{S_E}\brc{\der_4(S_E\cW_{5\alp})
 -\der_4 S_E\cW_{5\alp}} \nonumber\\
 \eql -4\der_4\cW_{5\alp} 
 = -8\der_4\bar{D}^2D_\alp X_5. 
\eea
This can be solved as
\be
 V_E\Phi_T = \der_4 V_5-\Sgm_5-\bar{\Sgm}_5, 
\ee
where $V_5\equiv -8X_5$,\footnote{
Thus, $\cW_{5\alp}$ is expressed as 
$\cW_{5\alp}=-\frac{1}{4}\bar{D}^2D_\alp V_5$. 
} 
and $\Sgm_5$ is a chiral superfield. 
Substituting this into (\ref{5DcL_VT:1}), we obtain 
\bea
 \cL_{\rm VT}^{\rm (5D)} \eql 
 \int\dr^4\tht\;f_{IJ}\left[\brc{
 -2\hat{\Sgm}^ID^\alp V^J\cW_{5\alp}
 +\frac{1}{2}\brkt{\der_4 V^ID^\alp V^J-\der_4 D^\alp V^IV^J}\cW_{5\alp}+\hc} 
 \right.\nonumber\\
 &&\hspace{15mm}
 +\brkt{\der_4 V_5-\Sgm_5-\bar{\Sgm}_5}\brkt{D^\alp V^I\cW_\alp^J
 +\bar{D}_{\dalp}V^I\bar{\cW}^{J\dalp}
 +V^ID^\alp\cW_\alp^J} \nonumber\\
 &&\hspace{15mm}\left.
 +\frac{2\brkt{\der_4 V_5-\Sgm_5-\bar{\Sgm}_5}}{\hat{V}_E^2}
 \brkt{\der_4 V^I-\hat{\Sgm}^I-\bar{\hat{\Sgm}}^I}
 \brkt{\der_4 V^J-\hat{\Sgm}^J-\bar{\hat{\Sgm}}^J}\right]. 
\eea
Notice that the ``shape-modulus'' superfield~$S_E$ 
completely disappears from the Lagrangian by the field redefinition~(\ref{5Dsf}). 

Since it follows that
\bea
 &&\brkt{\der_4 V_5-\Sgm_5-\bar{\Sgm}_5}\brkt{D^\alp V^I\cW_\alp^J
 +\bar{D}_{\dalp}V^I\bar{\cW}^{J\dalp}
 +V^ID^\alp\cW_\alp^J} \nonumber\\
 \eql \der_4 V_5D^\alp V^I\cW_\alp^J
 +\frac{1}{2}\der_4 V_5V^ID^\alp\cW^J_\alp
 -\Sgm_5\brkt{D^\alp V^I\cW_\alp^J+\bar{D}_{\dalp}V^I\bar{\cW}^{J\dalp}
 +V^ID^\alp\cW_\alp^J}+\hc \nonumber\\
 \eql \der_4 V_5D^\alp V^I\cW^J_\alp
 -\frac{1}{2}D^\alp\brkt{\der_4 V_5V^I}\cW^J_\alp
 -\Sgm_5D^\alp V^I\cW^J_\alp \nonumber\\
 &&+\Sgm_5V^I\bar{D}_{\dalp}\bar{\cW}^{J\dalp}
 -\Sgm_5V^ID^\alp\cW_\alp^J+\hc \nonumber\\
 \eql \frac{1}{2}\brkt{\der_4 V_5D^\alp V^I-\der_4D^\alp V_5V^I}\cW^J_\alp
 -\Sgm_5D^\alp V^I\cW^J_\alp+\hc, 
\eea
the above Lagrangian is rewritten as
\bea
 \cL_{\rm VT}^{\rm (5D)} \eql 
 \int\dr^4\tht\;f_{IJ}\left[\left\{
 -2\hat{\Sgm}^ID^\alp V^J\cW_{5\alp}
 +\frac{1}{2}\brkt{\der_4 V^ID^\alp V^J-\der_4 D^\alp V^IV^J}\cW_{5\alp}
 \right.\right.\nonumber\\
 &&\hspace{17mm}\left.
 -\Sgm_5D^\alp V^I\cW^J_\alp
 +\frac{1}{2}\brkt{\der_4 V_5D^\alp V^I
 -\der_4 D^\alp V_5V^I}\cW^J_\alp 
 +\hc\right\} \nonumber\\
 &&\hspace{17mm}\left.
 +\frac{2\brkt{\der_4 V_5-\Sgm_5-\bar{\Sgm}_5}}{\hat{V}_E^2}
 \brkt{\der_4 V^I-\hat{\Sgm}^I-\bar{\hat{\Sgm}}^I}
 \brkt{\der_4 V^J-\hat{\Sgm}^J-\bar{\hat{\Sgm}}^J}\right]. \nonumber\\
 \label{5DcL:local}
\eea
As shown in Appendix~\ref{derivation}, we find that 
\bea
 &&f_{IJ}\brc{
 \brkt{\der_4 V^ID^\alp V^J-\der_4 D^\alp V^IV^J}\cW_{5\alp}
 +\brkt{\der_4 V_5D^\alp V^I-\der_4D^\alp V_5V^I}\cW_\alp^J}+\hc 
 \nonumber\\
 \eql 2f_{IJ}
 \brkt{\der_4 V^I D^\alp V_5-\der_4 D^\alp V^I V_5}\cW^J_\alp
 +\hc.  \label{fml:dimred}
\eea
By means of this relation, (\ref{5DcL:local}) is further rewritten as 
\bea
 \cL_{\rm VT}^{\rm (5D)} \eql 
 \int\dr^4\tht\;f_{IJ}\left[\left\{
 -2\hat{\Sgm}^ID^\alp V^J\cW_{5\alp}
 -\Sgm_5D^\alp V^I\cW^J_\alp
 +\frac{1}{3}\brkt{\der_4 V^ID^\alp V^J-\der_4 D^\alp V^IV^J}\cW_{5\alp}
 \right.\right.\nonumber\\
 &&\hspace{9mm}\left.
 +\frac{1}{3}\brkt{\der_4 V_5D^\alp V^I
 -\der_4 D^\alp V_5V^I}\cW^J_\alp 
 +\frac{1}{3}\brkt{\der_4 V^ID^\alp V_5
 -\der_4D^\alp V^IV_5}\cW^J_\alp
 +\hc\right\} \nonumber\\
 &&\hspace{9mm}\left.
 +\frac{2\brkt{\der_4 V_5-\Sgm_5-\bar{\Sgm}_5}}{\hat{V}_E^2}
 \brkt{\der_4 V^I-\hat{\Sgm}^I-\bar{\hat{\Sgm}}^I}
 \brkt{\der_4 V^J-\hat{\Sgm}^J-\bar{\hat{\Sgm}}^J}\right]. 
 \label{5DcL:local:2}
\eea
Here we relabel $(V_5,\Sgm_5)$ as $(V^0,\Sgm^0)$. 
Then this Lagrangian is expressed as
\bea
 \cL_{\rm VT}^{\rm (5D)} \eql 
 \sbk{-\int\dr^2\tht\;C_{IJK}\Sgm^I\cW^J\cW^K+\hc} \nonumber\\
 &&+\int\dr^4\tht\;\frac{C_{IJK}}{3}
 \brc{\brkt{\der_4 V^I D^\alp V^J-\der_4 D^\alp V^IV^J}\cW^K_\alp+\hc} 
 \nonumber\\
 &&+\int\dr^4\tht\;\frac{2C_{IJK}}{3}\cV^I\cV^J\cV^K, 
 \label{5DcL_vec}
\eea
where the indices~$I,J,K$ now run from 0, 
the completely symmetric constant tensor~$C_{IJK}$ is defined as
$C_{IJ0} = f_{IJ}$ ($I,J\neq 0$) and the other components are zero, 
and 
\be
 \cV^I \equiv \der_4 V^I-\Sgm^I-\bar{\Sgm}^I, 
\ee
which is the extra-dimensional component of the field strength superfield. 

The 5D Lagrangians~(\ref{5DcL_hyp}) and (\ref{5DcL_vec}) 
perfectly agree with the $\cN=1$ superfield description of 5D SUGRA 
derived in Refs.~\cite{Paccetti:2004ri,Abe:2004ar}.

\section{Summary}  \label{summary}
We have found the $\cN=1$ superfield description of 6D SUGRA, 
and clarified how the moduli superfields appear in the action. 
We identified the combinations of the bosonic component fields that 
form $\cN=1$ superfields. 
By acting the SUSY transformations on them, we can identify 
the fermionic components of the superfields, 
which are expected to have complicated forms. 
Our result~(\ref{cL:SUGRA}) reproduces the action in the global SUSY case 
by replacing the moduli superfields~$V_E$ and $S_E$ 
with their constant background values. 
We have also shown that it is gauge-invariant both 
under (\ref{gauge_trf:tensor}) and (\ref{gauge_trf:local}), 
and is consistent with the known superfield action of 5D SUGRA 
through the dimensional reduction. 

Compared to 5D SUGRA, 
the existence of the tensor multiplet and the ``shape'' modulus~$S_E$ make 
the construction of the action complicated. 
In the global SUSY limit, the tensor multiplet is described by 
on-shell superfields that are subject to the constraints in (\ref{constraints}). 
When the theory is promoted to SUGRA, this multiplet becomes off-shell 
and the superfields~$X_4$ (or $X_5$) and $Y_\alp$ can be treated 
as unconstrained independent superfields. 
As shown in Sec.~\ref{gauge_inv}, the gauge invariance of the action 
in the vector-tensor sector is realized in a quite nontrivial manner 
because the Lagrangian is invariant only up to total derivatives. 
The gauge invariance strictly restricts the $S_E$-dependence of the action. 
It appears in the action through $\der_E$ and $U_E(S_E,\bar{S}_E)$ 
defined in (\ref{def:der_E}) and (\ref{def:U_E}), respectively. 
We should also note that the $S_E$-dependence is absorbed by the field redefinition 
and completely disappears when one of the extra dimensions is reduced. 
This is another nontrivial check for our result. 

In this work, we have neglected the fluctuation modes of $e_\mu^{\;\;\udl{\nu}}$, 
$e_\mu^{\;\;\udl{n}}$ and $e_m^{\;\;\udl{\nu}}$ 
($\mu,\nu=0,1,2,3$; $m,n=4,5$). 
As mentioned in the footnote~\ref{FN:4Dgravity}, 
the fluctuations of $e_\mu^{\;\;\udl{\nu}}$ can be taken into account 
by using the invariant action formulae in the superconformal formulation 
of 4D SUGRA. 
As for the ``off-diagonal'' components~$e_\mu^{\;\;\udl{n}}$ and $e_m^{\;\;\udl{\nu}}$, 
further effort is necessary. 
However, we expect that it is not very difficult to incorporate them at linear order 
by means of the linearized SUGRA 
formulation~\cite{Ferrara:1977mv,Siegel:1978mj,Sakamura:2011df}, 
just like the 5D SUGRA case discussed in Refs.~\cite{Linch:2002wg,Sakamura:2012bj}. 

Our superfield description is useful to derive 4D effective theories of 
various 6D SUGRA models, 
as we did in the 5D SUGRA case~\cite{Abe:2006eg,Abe:2008an,Abe:2011rg}. 
Especially, we can treat 
a case that there exists the background magnetic flux penetrating 
the compact space or that the compact space has nonvanishing curvature. 
An explicit derivation of 4D effective theory will be discussed 
in a subsequent paper.

\subsection*{Acknowledgements}
H.A., Y.S. and Y.Y. are supported in part by Grant-in-Aid for Young Scientists (B) 
(No. 25800158),
Grant-in-Aid for Scientific Research (C) (No.25400283),    
and Research Fellowships for Young Scientists (No.26-4236), respectively, 
which are from Japan Society for the Promotion of Science.

\appendix

\section{6D and 4D superconformal algebras} \label{SCalgebras}
The 6D superconformal algebra consists of 
the translation~$P_A$ ($A=0,1,\cdots,5$), 
the local Lorentz transformation~$M_{AB}$, 
the dilatation~$D$, the special conformal transformation~$K_A$, 
the $\suU$ automorphism~$U^{ij}$, 
SUSY~$Q_{\ualp}^i$ and the conformal SUSY~$S_{\ualp}^i$.\footnote{
Note that $Q_{\ualp}^i$ and $S^{i\ualp}$ are $\suU$-Majorana-Weyl spinors. 
We follow the notation of Ref.~\cite{Abe:2015bqa} for 6D spinors.  
} 
Here, $\ualp=1,2,3,4$ is the  6D Weyl spinor index, 
and $i=1,2$ is the $\suU$-doublet index.  
They satisfy the following algebra. 
\bea
 \sbk{M_{AB},M_{CD}} \eql i\brkt{\eta_{BC}M_{AD}-\eta_{AC}M_{BD}
 -\eta_{BD}M_{AC}+\eta_{AD}M_{BC}}, \nonumber\\
 \sbk{M_{AB},P_C} \eql i\brkt{\eta_{BC}P_A-\eta_{AC}P_B}, \nonumber\\
 \sbk{M_{AB},K_C} \eql i\brkt{\eta_{BC}K_A-\eta_{AC}K_B}, \nonumber\\
 \sbk{M_{AB},D} \eql 0, \;\;\;\;\;
 \sbk{D,P_A} = iP_A, \;\;\;\;\;
 \sbk{D,K_A} = -iK_A, \nonumber\\
 \sbk{P_A,K_B} \eql 2i\brkt{\eta_{AB}D+M_{AB}}, 
\eea
and 
\bea
 \sbk{M_{AB},Q_{\ualp}^i} \eql \frac{i}{2}\brkt{\gm_{AB}Q^i}_{\ualp}, \;\;\;\;\;
 \sbk{D,Q_{\ualp}^i} = \frac{i}{2}Q_{\ualp}^i, \nonumber\\
 \sbk{P_A,Q_{\ualp}^i} \eql 0, \;\;\;\;\;
 \sbk{K_A,Q_{\ualp}^i} = (\gm_A S^i)_{\ualp}, \nonumber\\
 \sbk{M_{AB},S^{i\ualp}} \eql \frac{i}{2}\brkt{\tl{\gm}_{AB}S^i}^{\ualp}, \;\;\;\;\;
 \sbk{D,S^{i\ualp}} = -\frac{i}{2}S^{i\ualp}, \nonumber\\
 \sbk{P_A,S^{i\ualp}} \eql \brkt{\tl{\gm}_AQ^i}^{\ualp}, \;\;\;\;\;
 \sbk{K_A,S^{i\ualp}} = 0, \nonumber\\
 \brc{Q_{\ualp}^1,Q_{\ubt}^2} \eql 2\brkt{\gm^A C^{-1}}_{\ualp\ubt}P_A, \nonumber\\
 \brc{Q_{\ualp}^i,S^{j\ubt}} \eql -i\ep^{ij}
 \brc{\brkt{\gm^{AB}\tl{C}^{-1}}_{\ualp}^{\;\;\ubt}M_{AB}
 -2\brkt{\tl{C}^{-1}}_{\ualp}^{\;\;\ubt}D}
 +8\brkt{\tl{C}^{-1}}_{\ualp}^{\;\;\ubt}U^{ij}, \nonumber\\
 \brc{S^{1\ualp},S^{2\ubt}} \eql 2\brkt{\tl{\gm}^A\tl{C}^{-1}}^{\ualp\ubt}K_A, 
 \nonumber\\
 \sbk{U^{ij},U^{kl}} \eql \ep^{li}U^{kj}-\ep^{jk}U^{il}, \nonumber\\
 \sbk{U^{ij},Q_{\ualp}^k} \eql -\ep^{jk}Q_{\ualp}^i
 -\frac{1}{2}\ep^{ij}Q_{\ualp}^k, \;\;\;\;\;
 \sbk{U^{ij},S^{k\ualp}} = -\ep^{jk}S^{i\ualp}
 -\frac{1}{2}\ep^{ij}S^{k\ualp}. 
 \label{6DSCalgebra:2}
\eea
Here we decompose the 4-component spinors into 2-component ones as
\bea
 Q_{\ualp}^1 \eql \begin{pmatrix} Q^1_\alp \\ -\bar{Q}^{2\dalp} \end{pmatrix}, 
 \;\;\;\;\;
 Q_{\ualp}^2 = \begin{pmatrix} Q^2_\alp \\ \bar{Q}^{1\dalp} \end{pmatrix}, 
 \nonumber\\
 S^{1\ualp} \eql \begin{pmatrix} S^{1\alp} \\ -\bar{S}^2_{\dalp} \end{pmatrix}, 
 \;\;\;\;\;
 S^{2\ualp} = \begin{pmatrix} S^{2\alp} \\ \bar{S}^1_{\dalp} \end{pmatrix}. 
\eea
The $\suU$ generators~$U^{ij}$ are also expressed as
\be
 U^i_{\;\;j} = \ep_{jk}U^{ik} = \sum_{a=1}^3u^a(\sgm^a)^i_{\;\;j}. 
\ee
From (\ref{6DSCalgebra:2}), we obtain 
\bea
 \sbk{M_{\mu\nu},Q^1_\alp} \eql i\brkt{\sgm^{\mu\nu}Q^1}_\alp, \;\;\;\;\;
 \sbk{M_{\mu\nu},S^2_\alp} = i\brkt{\sgm^{\mu\nu}S^2}_\alp, \nonumber\\
 \sbk{M_{45},Q^1_\alp} \eql -\frac{1}{2}Q^1_\alp, \;\;\;\;\;
 \sbk{M_{45},S^2_\alp} = \frac{1}{2}S^2_\alp, \nonumber\\
 \sbk{D,Q^1_\alp} \eql \frac{i}{2}Q^1_\alp, \;\;\;\;\;
 \sbk{D,S^2_\alp} = -\frac{i}{2}S^2_\alp, \nonumber\\
 \sbk{K_\mu,Q^1_\alp} \eql \brkt{\sgm_\mu\bar{S}^2}_\alp, \;\;\;\;\;
 \sbk{P_\mu,S^2_\alp} = \brkt{\sgm_\mu\bar{Q}^1}_\alp, \nonumber\\
 \brc{Q_\alp^1,\bar{Q}_{\dbt}^1} \eql -2\sgm^\mu_{\alp\dbt}P_\mu, \;\;\;\;\;
 \brc{S^2_\alp,\bar{S}^2_{\dbt}} = -2\sgm^\mu_{\alp\dbt}K_\mu, \nonumber\\
 \brc{Q_\alp^1,S^{2\bt}} \eql 2i(\sgm^{\mu\nu})_\alp^{\;\;\bt}M_{\mu\nu}
 -2\dlt_\alp^{\;\;\bt}\brkt{M_{45}-4u^3+iD},  \nonumber\\
 \sbk{u^3,Q^1_\alp} \eql -\frac{1}{2}Q^1_\alp, \;\;\;\;\;
 \sbk{u^3,S^2_\alp} = \frac{1}{2}S^1_\alp, 
\eea
in the 2-component-spinor notation. 
This is the 4D $\cN=1$ superconformal algebra, and 
we can identify the generator of the U(1)$_A$ automorphism as
\be
 \cQ_A = M_{45}-4u^3. \label{id:cQ_A}
\ee
We have normalized $\cQ_A$ so that $Q^1_\alp$ and $S^2_\alp$ have 
the charges $3/2$ and $-3/2$, respectively.

\section{SUSY transformation of 6D Weyl multiplet} \label{trf:Weyl}
The 6D Weyl multiplet consists of the sechsbein~$e_M^{\;\;\udl{N}}$, 
the gravitino~$\Psi^i_{M\ualp}$, 
the gauge fields for the dilatation~$b_M$ and for the $\suU$ automorphism~$V_M^a$ 
($a=1,2,3$), the anti-self-dual tensor~$T^-_{\udl{M}\udl{N}\udl{L}}$, 
and some auxiliary fields. 
The SUSY transformations of the (extra-dimensional-components of) 
6D Weyl multiplet~\cite{Kugo:2000hn,Bergshoeff:1985mz} are expressed 
in the 2-component spinor notation as follows.\footnote{ 
Since we neglect the fluctuations of $e_\mu^{\;\;\udl{\nu}}$, 
$e_\mu^{\;\;\udl{n}}$ and $e_m^{\;\;\udl{\nu}}$, 
we do not discriminate the curved indices from the flat ones for the 4D part. 
}
\bea
 \dlt_\ep e_m^{\;\;\udl{4}} \eql 
 2\brkt{\ep^1\psi_m^2-\ep^2\psi_m^1}+\hc, \nonumber\\
 \dlt_\ep e_m^{\;\;\udl{5}} \eql 
 -2i\brkt{\ep^1\psi_m^2-\ep^2\psi^1_m}+\hc, \nonumber\\
 \dlt_\ep\psi_m^1 \eql \left\{\der_m+\frac{1}{2}b_m
 -\frac{1}{2}\brkt{\omg_m^{\;\;\mu\nu}\sgm_{\mu\nu}
 +i\omg_m^{\;\;\udl{4}\udl{5}}}-iV_m^3 
 +\frac{e_m^{\;\;\udl{4}}-ie_m^{\;\;\udl{5}}}{4}
 \brkt{T_{\mu\nu\udl{4}}+iT_{\mu\nu\udl{5}}}\sgm^{\mu\nu}\right\}\ep^1 \nonumber\\
 &&-i\brkt{V_m^1-iV_m^2}\ep^2 \nonumber\\
 &&+\brc{\frac{i}{2}\brkt{\omg_m^{\;\;\mu\udl{4}}
 +i\omg_m^{\;\;\mu\udl{5}}}\sgm_\mu
 -\frac{e_m^{\;\;\udl{4}}+ie_m^{\;\;\udl{5}}}{24}
 \ep^{\mu\nu\rho\lmd}T^-_{\mu\nu\rho}\sgm_\lmd
 +6T^-_{\mu\udl{4}\udl{5}}\sgm^\mu}\bar{\ep}^2,  \nonumber\\
 \dlt_\ep\psi_m^2 \eql \left\{\der_m+\frac{1}{2}b_m
 -\frac{1}{2}\brkt{\omg_m^{\;\;\mu\nu}\sgm_{\mu\nu}+i\omg_m^{\;\;\udl{4}\udl{5}}}
 +iV_m^3
 +\frac{e_m^{\;\;\udl{4}}-ie_m^{\;\;\udl{5}}}{4}
 \brkt{T_{\mu\nu\udl{4}}+iT_{\mu\nu\udl{5}}}\sgm^{\mu\nu}\right\}\ep^2 
 \nonumber\\
 &&-\left\{\frac{i}{2}
 \brkt{\omg_m^{\;\;\mu\udl{4}}+i\omg_m^{\;\;\mu\udl{5}}}\sgm_\mu
 -\frac{e_m^{\;\;\udl{4}}+ie_m^{\;\;\udl{5}}}{24}
 \brkt{\ep^{\mu\nu\rho\lmd}T^-_{\mu\nu\rho}\sgm_\lmd+6T^-_{\mu\udl{4}\udl{5}}\sgm^\mu}\right\}
 \bar{\ep}^1 \nonumber\\
 &&-i\brkt{V_m^1+iV_m^2}\ep^1, \nonumber\\
 &\vdots & \label{SUSYtrf:Weyl}
\eea
where the 2-component spinors~$\psi_m^i$ $(i=1,2)$ are embedded into 
the 4-component ones as
\be
 \Psi_{m\ualp}^1 = \begin{pmatrix} \psi^1_{m\alp} \\ -\bar{\psi}^{2\dalp}_m \end{pmatrix}, 
 \;\;\;\;\;
 \Psi_{m\ualp}^2 = \begin{pmatrix} \psi^2_{m\alp} \\ \bar{\psi}^{1\dalp}_m \end{pmatrix}, 
\ee
which have positive 6D chiralities. 
In Sec.~\ref{moduli_sf}, we focus on a half of the whole SUSY 
parameterized by $\ep^1_\alp$ and $\bar{\ep}^1_{\dalp}$.

\section{Component expression of constraint~(\ref{rel:XYZ:local})} \label{comp_of_constraint}
Here we express the constraint~(\ref{rel:XYZ:local}) in terms of 
the component fields, and clarify the independent degrees of freedom. 
Note that (\ref{rel:XYZ:local}) is rewritten as
\be
 \bar{D}^2\brkt{D_\alp X_5+4\der_5 Y_\alp} 
 = \frac{1}{S_E^2}\bar{D}^2\brkt{D_\alp X_4+4\der_4 Y_\alp}. 
\ee
Since $\bar{D}^2D_\alp X_m$ ($m=4,5$) are field strength superfields, 
$4\der_m\bar{D}^2 Y_\alp$ are chiral spinor superfields 
and $1/S_E^2$ is a chiral scalar superfield, 
they are expanded as
\bea
 \bar{D}^2D_\alp X_m \eql \lmd_{m\alp}+\tht_\alp D_m
 +i(\sgm^{\mu\nu}\tht)_\alp v_{m\mu\nu}-i\tht^2(\sgm^\mu\der_\mu\bar{\lmd}_m)_\alp, 
 \nonumber\\
 4\der_m\bar{D}^2Y_\alp \eql \omg_{m\alp}+\tht_\alp K_m
 +i(\sgm^{\mu\nu}\tht)_\alp K_{m\mu\nu}+\tht^2\tau_{m\alp}, \nonumber\\
 \frac{1}{S_E^2} \eql a+\tht\psi+\tht^2F, 
\eea
where $D_m$ is a real scalar, 
$v_{m\mu\nu}\equiv\der_\mu v_{m\nu}-\der_\nu v_{m\mu}$ is a field strength, 
$K_m$ is a complex scalar and $K_{m\mu\nu}$ is a real antisymmetric tensor. 
Then, we calculate
\bea
 4\der_5\bar{D}^2Y_\alp \eql \frac{1}{S_E^2}\bar{D}^2\brkt{D_\alp X_4+4\der_4 Y_\alp}
 -\bar{D}^2D_\alp X_5 \nonumber\\
 \eql a\brkt{\lmd_4+\omg_4}_\alp-\lmd_{5\alp} \nonumber\\
 &&+\tht_\alp\brc{a\brkt{D_4+K_4+\frac{1}{2}\psi(\lmd_4+\omg_4)}-D_5} \nonumber\\
 &&+i(\sgm^{\mu\nu}\tht)_\alp
 \brkt{\frac{1}{2}\ep_{\mu\nu\rho\lmd}C_{\rm 4R}^{\rho\lmd}
 +C_{4{\rm I}\mu\nu}-v_{5\mu\nu}} \nonumber\\
 &&+\tht^2\left\{
 F(\lmd_4+\omg_4)_\alp-\frac{1}{2}\psi_\alp(D_4+K_4)
 -\frac{i}{2}(\sgm^{\mu\nu}\psi)_\alp (v_{4\mu\nu}+K_{4\mu\nu}) \right.\nonumber\\
 &&\hspace{10mm}\left.
 +a\brkt{\tau_4-i\sgm^\mu\der_\mu\bar{\lmd}_4}_\alp+i(\sgm^\mu\der_\mu\bar{\lmd}_5)_\alp
 \right\},  \label{comp:constraint}
\eea
where 
\bea
 C_{4{\rm R}\mu\nu} \defa (\Re a)\brkt{v_{4\mu\nu}+K_{4\mu\nu}}
 -\Re\brc{\frac{a}{2}\psi\sgm_{\mu\nu}\brkt{\lmd_4+\omg_4}}, \nonumber\\
 C_{4{\rm I}\mu\nu} \defa (\Im a)\brkt{v_{4\mu\nu}+K_{4\mu\nu}}
 -\Im\brc{\frac{a}{2}\psi\sgm_{\mu\nu}\brkt{\lmd_4+\omg_4}}. 
\eea
We have used that
\bea
 (\tht\psi)\tl{\lmd}_\alp \eql \frac{1}{2}\brc{(\psi\tl{\lmd})\tht_\alp
 -(\psi\sgm^{\mu\nu}\tl{\lmd})(\sgm_{\mu\nu}\tht)_\alp}, \nonumber\\
 \brkt{C_{4{\rm R}\mu\nu}+iC_{4{\rm I}\mu\nu}}(\sgm^{\mu\nu}\tht)_\alp 
 \eql i\brkt{\frac{1}{2}\ep_{\mu\nu\rho\lmd}C_{4{\rm R}}^{\rho\lmd}
 +C_{4{\rm I}\mu\nu}}(\sgm^{\mu\nu}\tht)_\alp, 
\eea
where $\tl{\lmd}_\alp\equiv\lmd_{4\alp}+\omg_{4\alp}$. 
 
From (\ref{comp:constraint}), we can see that
the constraint~(\ref{rel:XYZ:local}) can be satisfied for a given $X_4$ and $Y_\alp$ 
by adjusting $X_5$ and $S_E$. 
Specifically, for given values of $\bar{D}^2D_\alp X_4$ and $4\der_4\bar{D}^2Y_\alp$, 
we can realize any values for $\omg_{5\alp}$, $K_5$, $K_{5\mu\nu}$ and $\tau_{5\alp}$ 
in $4\der_5\bar{D}^2Y_\alp$ by tuning $\lmd_{5\alp}$, $D_5$ and $a$,
$v_{5\mu}$ and two real degrees of freedom in $\psi_\alp$, 
and $F$ and the remaining degrees of freedom in $\psi_\alp$, respectively.

\section{Derivation of Eq.(\ref{fml:dimred})} \label{derivation}
Here we derive the relation~(\ref{fml:dimred}). 
We neglect total derivatives. 
Then we obtain 
\bea
 A \defa f_{IJ}\brkt{\der_4V_5D^\alp V^I
 -\der_4 D^\alp V_5V^I}\cW^J_\alp+\hc 
 \nonumber\\
 \eql -f_{IJ}\brkt{V_5\der_4 D^\alp V^I
 -D^\alp V_5\der_4 V^I}\cW^J_\alp+B+\hc,  \label{def:A}
\eea
where
\be
 B \equiv -f_{IJ}\brkt{V_5D^\alp V^I
 -D^\alp V_5V^I}\der_4\cW^J_\alp. 
\ee
We can show that
\bea
 B+\hc \eql \frac{f_{IJ}}{4}\bar{D}^2
 \brkt{V_5D^\alp V^I-D^\alp V_5V^I}
 \der_4D_\alp V^J+\hc \nonumber\\
 \eql f_{IJ}\cW^\alp_5V^I\der_4 D_\alp V^J+C+\hc, 
\eea
where
\bea
 C \defa \frac{f_{IJ}}{4}\left(
 \bar{D}^2V_5D^\alp V^I
 +2\bar{D}_{\dalp}V_5\bar{D}^{\dalp}D^\alp V^I
 +V_5\bar{D}^2D^\alp V^I \right.\nonumber\\
 &&\hspace{10mm}\left.
 +2\bar{D}_{\dalp}D^\alp V_5\bar{D}^{\dalp}V^I
 -D^\alp V_5\bar{D}^2V^I\right)\der_4 D_\alp V^J. 
\eea
Here, it follows that
\bea
 C+\hc \eql -\frac{f_{IJ}}{4}D^\alp\left(
 \bar{D}^2V_5D_\alp V^I
 +2\bar{D}_{\dalp}V_5\bar{D}^{\dalp}D_\alp V^I
 +V_5\bar{D}^2D_\alp V^I \right.\nonumber\\
 &&\hspace{15mm}\left.
 +2\bar{D}_{\dalp}D_\alp V_5\bar{D}^{\dalp}V^I
 -D_\alp V_5\bar{D}^2V^I\right)\der_4 V^J+\hc \nonumber\\
 \eql -\frac{f_{IJ}}{4}\left(
 D^\alp\bar{D}^2V_5D_\alp V^I
 -2\bar{D}_{\dalp}V_5D^\alp\bar{D}^{\dalp}D_\alp V^I
 +D^\alp V_5\bar{D}^2D_\alp V^I \right.\nonumber\\
 &&\hspace{10mm}\left.
 +V_5D^\alp\bar{D}^2D_\alp V^I
 +2D^\alp\bar{D}_{\dalp}D_\alp V_5D_\alp V^I
 +D_\alp V_5D^\alp\bar{D}^2V^I\right)\der_4 V^J+\hc \nonumber\\
 \eql -\frac{f_{IJ}}{4}\left(
 \bar{D}^2D^\alp V_5D_\alp V^I
 +4i\sgm^\mu_{\alp\dalp}\der_\mu\bar{D}^{\dalp}V_5D^\alp V^I
 +2\bar{D}_{\dalp}V_5D^2\bar{D}^{\dalp}V^I \right.\nonumber\\
 &&\hspace{10mm}
 -4i\sgm^\mu_{\alp\dalp}\bar{D}^{\dalp}V_5\der_\mu D^\alp V^I
 +V_5D^\alp\bar{D}^2D_\alp V^I
 -2D^2\bar{D}_{\dalp}V_5\bar{D}^{\dalp}V^I \nonumber\\
 &&\hspace{10mm}\left. 
 -4i\sgm^\mu_{\alp\dalp}\der_\mu D^\alp V_5\bar{D}^{\dalp}V^I
 +4i\sgm^\mu_{\alp\dalp}D^\alp V_5\der_\mu\bar{D}^{\dalp}V^I\right)
 \der_4 V^J+\hc \nonumber\\
 \eql f_{IJ}\brkt{-\cW^\alp_5D_\alp V^I
 +2\bar{D}_{\dalp}V_5\bar{\cW}^{I\dalp}
 +V_5D^\alp\cW_\alp^I}\der_4 V^J+\hc \nonumber\\
 \eql f_{IJ}\sbk{
 -D^\alp V^I\der_4 V^J\cW_{5\alp}
 +\brc{2D^\alp V_5\der_4 V^I
 -D^\alp\brkt{V_5\der_4 V^I}}\cW^J_\alp}+\hc \nonumber\\
 \eql f_{IJ}\brc{-D^\alp V^I\der_4 V^J\cW_{5\alp}
 +\brkt{D^\alp V_5\der_4 V^I-V_5\der_4 D^\alp V^I}\cW^J_\alp}+\hc. 
\eea
We have used the commutation relations:
\be
 \brc{D_\alp,\bar{D}_{\dalp}} = -2i\sgm^\mu_{\alp\dalp}\der_\mu, 
 \;\;\;\;\;
 \sbk{D_\alp,\bar{D}^2} = -4i\sgm^\mu_{\alp\dalp}\der_\mu\bar{D}^{\dalp}. 
\ee
Therefore, (\ref{def:A}) is calculated as
\bea
 A \eql -f_{IJ}\brkt{V_5\der_4 D^\alp V^I
 -D^\alp V_5\der_4 V^I}\cW^J_\alp
 +f_{IJ}\cW^\alp_5V^I\der_4 D_\alp V^J \nonumber\\
 &&-f_{IJ}D^\alp V^I\der_4 V^J\cW_{5\alp}
 +f_{IJ}\brkt{D^\alp V_5\der_4 V^I-V_5\der_4D^\alp V^I}
 \cW^J_\alp+\hc \nonumber\\
 \eql 2f_{IJ}\brkt{\der_4 V^I D^\alp V_5
 -\der_4 D^\alp V^I V_5}\cW^J_\alp \nonumber\\
 &&-f_{IJ}\brkt{\der_4 V^ID^\alp V^J-\der_4 D^\alp V^IV^J}\cW_{5\alp}
 +\hc. 
\eea
Namely, we obtain 
\bea
 &&f_{IJ}\brc{\brkt{\der_4 V^I D^\alp V^J-\der_4 D^\alp V^I V^J}\cW_{5\alp}
 +\brkt{\der_4 V_5D^\alp V^I-\der_4 D^\alp V_5V^I}\cW^J_\alp}+\hc 
 \nonumber\\
 \eql 2f_{IJ}\brkt{\der_4 V^ID^\alp V_5
 -\der_4D^\alp V^IV_5}\cW^J_\alp+\hc. 
\eea


\end{document}